\documentclass[pra,nofootinbib,twocolumn,superscriptaddress,showpacs]{revtex4}

\pdfoutput=1

\usepackage{amsmath,amsfonts,amssymb,bm}
\usepackage{dcolumn}
\usepackage[final]{graphicx}
\usepackage{bm}
\usepackage{comment}

\newcommand{\be}{\begin{equation}}
\newcommand{\ee}{\end{equation}}
\newcommand{\bea}{\begin{eqnarray}}
\newcommand{\eea}{\end{eqnarray}}
\newcommand{\bit}{\begin{itemize}}
\newcommand{\eit}{\end{itemize}}

\newcommand{\bfl}{\begin{flushright}}
\newcommand{\efl}{\end{flushright}}

\newcommand{\non}{\nonumber \\}

\newcommand{\ra}{\rangle}
\newcommand{\la}{\langle}

\newcommand{\ah}{\hat{a}}

\includecomment{pdffigure}

\begin{document}

\bibliographystyle{apsrev}

\title{Optimal control of number squeezing in trapped Bose-Einstein condensates}

\author{Julian Grond}
\affiliation{Institut f\"ur Physik,
  Karl--Franzens--Universit\"at Graz, Universit\"atsplatz 5,
  8010 Graz, Austria}

\author{Gregory von Winckel}
\affiliation{Institut f\"ur Mathematik und Wissenschaftliches Rechnen, 
Karl-Franzens-Universit\"at Graz, Heinrichstra\ss e 36,
  8010 Graz, Austria}

\author{J\"org Schmiedmayer}
\affiliation{Atominstitut der \"Osterreichischen Universit\"aten,
  TU--Wien, Stadionallee 2, 1020 Wien, Austria}
  
\author{Ulrich Hohenester}
\affiliation{Institut f\"ur Physik,
  Karl--Franzens--Universit\"at Graz, Universit\"atsplatz 5,
  8010 Graz, Austria}

\date{\today}

\begin{abstract}

We theoretically analyze atom interferometry based on trapped ultracold atoms, and employ optimal control theory in order to optimize number squeezing and condensate trapping. In our simulations, we consider a setup where the confinement potential is transformed from a single to a double well, which allows to split the condensate. To avoid in the ensuing phase-accumulation stage of the interferometer dephasing due to the nonlinear atom-atom interactions, the atom number fluctuations between the two wells should be sufficiently low. We show that low number fluctuations (high number squeezing) can be obtained by optimized splitting protocols. Two types of solutions are found: in the \emph{Josephson regime} we find an oscillatory tunnel control and a parametric amplification of number squeezing, while in the \emph{Fock regime} squeezing is obtained solely due to the nonlinear coupling, which is transformed to number squeezing by peaked tunnel pulses. We study splitting and squeezing within the frameworks of a generic two-mode model, which allows us to study the basic physical mechanisms, and the \emph{multi-configurational time dependent Hartree for bosons} method, which allows for a microscopic modeling of the splitting dynamics in realistic experiments. Both models give similar results, thus highlighting the general nature of these two solution schemes. We finally analyze our results in the context of atom interferometry.

\end{abstract}



\maketitle

\section{Introduction}
 
Squeezed states offer the possibility of measurements below shot noise \cite{giovannetti:04,wineland:94}. In matter wave interferometry \cite{cronin:09} with Bose Einstein condensates (BECs), number squeezing in the split BEC helps to slow down the phase diffusion arising from atom-atom interactions \cite{javanainen:97}, and thus the potential measurement time and with it the sensitivity of the atom interferometer is increased. 

Splitting a trapped BEC is achieved by transforming the confinement potential from a single to a double well \cite{shin:04,schumm:05}. Close to the splitting point, where the condensate breaks apart into two spatially separated parts, the competition between tunneling and nonlinear interaction leads to a decrease of number fluctuations and squeezing. 

In this paper we will discuss our concepts to optimize squeezing and interference in trapped BECs in the context of Atom chips \cite{folman:00,folman:02,fortagh:07} which provide a versatile tool for precise manipulation of trapped BECs and allow robust implementation of the splitting procedure and interference experiments \cite{schumm:05,wang:05}. A key ingredient in our consideration is the beam splitter for the trapped or guided atoms \cite{cassettari:00b,cassettari:00,haensel:01,lesanovsky:06}. As an example we will consider here the RF beam splitter \cite{lesanovsky:06,lesanovsky:06b,hofferberth:06} as employed in the atom chip experiments.  However, our considerations can be easily applied to any other double well splitting and trapped atom interference setup \cite{shin:04,albiez:05,wang:05,esteve:08}.

Squeezing in the context of BEC splitting has been frequently discussed by assuming an approximately exponential decrease of the tunnel coupling close to the splitting point \cite{javanainen:99,pezze:05,jaeaeskelaeinen:04,menotti:01}. In experiment, number squeezing has been achieved in an optical trap \cite{chuu:05}, in optical lattices \cite{orzel:01,greiner:02,gerbier:06,sebby:07,li:07}, and more recently through linear splitting \cite{jo:07,esteve:08}.  In Ref.~\cite{grond.praR:09} we have recently shown that exponential splitting is by far not optimal, and there exist alternative squeezing protocols allowing for much more squeezing. These solutions are obtained within the framework of optimal control theory (OCT) \cite{peirce:88,hohenester.pra:07}. In this paper we give details on our work about optimal number squeezing. 

In our simulations we go beyond the simple two-mode model and consider additionally the spatial dynamics, in order to have a realistic description of the splitting dynamics and to evaluate the importance of condensate excitations. Our theoretical approach is based on the \emph{multi-configurational time-dependent Hartree for bosons} (MCTDHB) method \cite{alon:08}, which allows to describe both atom number and spatial dynamics in a self-consistent fashion. This method allows us to use as a control parameter the splitting distance of the trap used in experiments. The generic two-mode model, whose control is through the tunnel coupling which is determined indirectly through the confinement potential, is used to obtain a simpler interpretation of the underlying control strategies. 

We find that there exist even simpler splitting protocols compared to our earlier work \cite{grond.praR:09}. These solutions can be understood similarly to the control schemes of Refs.~\cite{kitagawa:93,sorensen:01,law:01,jin:08} which rely on a constant, optimal value of the tunnel coupling. We compare the OCT results to a two-parameter optimization, where the control consists of an initial exponential splitting followed by a constant value of the tunnel coupling.  While within the generic two-mode model the results almost coincide, this simple strategy is not generally applicable within a realistic model for two main reasons: first, for short splitting times (on a time scale determined by the trapping potential) the condensates oscillate, and the split condensate ends up in an excited state; second, after number squeezing the condensates have to be further separated to turn off any tunnel coupling, which for realistic traps can only be achieved with refined protocols. To avoid these problems, we are seeking for control protocols where at the final time the condensates are at rest and fully decoupled. 

We have organized our paper as follows. In Sec.~II we give a brief description of our model system and its dynamic equations. Squeezing and squeezing optimization within a generic two-mode model is described in Sec.~III. In Sec.~IV we introduce our approach for the realistic simulation of condensate splitting, and discuss splitting and its optimizations within the MCTDHB framework. We show that the general trends of the two-mode model and MCTDHB are in agreement. Finally, in Sec.~V we analyze our optimized splitting procedures in the context of atom interferometry.

\section{Condensate splitting}

\subsection{Model}

\begin{figure}
\centering 
\includegraphics[width=0.8\columnwidth]{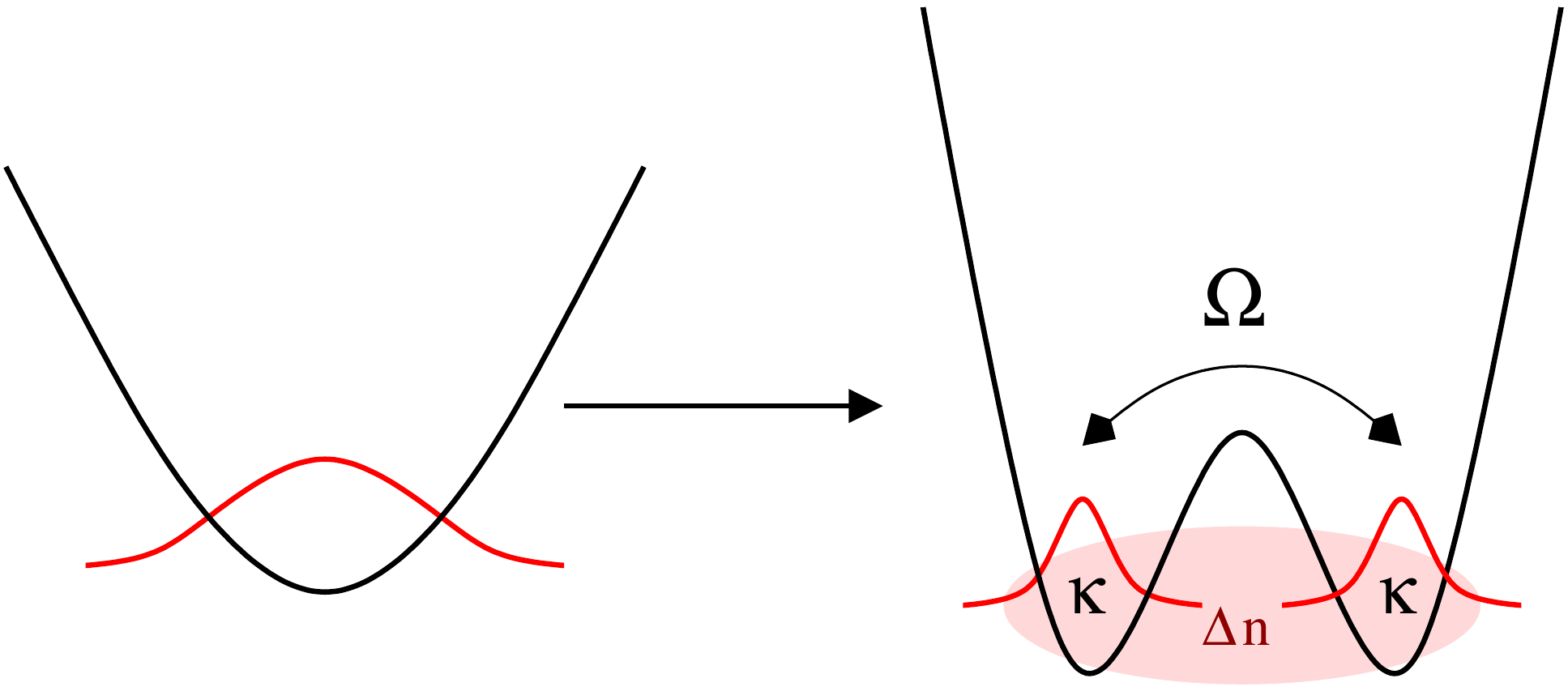}
\caption{ (Color online) Schematics of the splitting process. When splitting a condensate, it breaks up into two parts (red solid lines) with equal mean atom number, and number fluctuations $\Delta n$. The distribution of atoms between the wells is controlled by the tunnel coupling $\Omega$ and the nonlinear intra-well energy $\kappa$.\label{fig:1}}
\end{figure}

In our theoretical approach, we consider an elongated condensate whose wave function is modified only along a single spatial direction $x$ \cite{lesanovsky:06}. The atom dynamics can be described by the Hamiltonian \cite{dalfovo:99,leggett:01}
\begin{equation}\label{eq:ham}
  \hat H=\int\left[\hat\Psi^\dagger\hat h(x)
  \hat\Psi+\frac{U_0}{2}\hat\Psi^\dagger\hat\Psi^\dagger\hat\Psi\hat\Psi
  \right]\,dx\,.
\end{equation}
The first term on the right-hand side (rhs), which comprises the single particle or \emph{bare} Hamiltonian $\hat h(x)=-\frac 12\nabla^2+V_\lambda(x)$, accounts for the kinetic energy and the magnetic confinement potential $V_\lambda(x)$. The control parameter $\lambda(t)$ describes the variation of the confining potential when changing the external parameters, such as currents or rf-fields \cite{folman:02,lesanovsky:06}. Through $\lambda(t)$ it is possible to manipulate the trapped Bose-Einstein condensate, e.g. to split it by varying the potential from a single to a double well, see Fig.~\ref{fig:1}. The second term on the rhs accounts for the atom-atom interactions, which is scaled with the 1D-interaction parameter $U_0$. The field operators $\hat\Psi(x)$ and $\hat\Psi^\dagger(x)$ obey the usual equal-time commutation relations. 

We consider $^{87}$Rb atoms, and use units where $\hbar=1$, the mass of the atoms is set to one and length is measured in micrometers. The natural unit of time (energy) is then $1.37$ ms ($5.58$ nano Kelvin), and the results for the generic model can be arbitrarily scaled.

\subsection{Confinement potential}

For the realistic modeling of the splitting process, we consider the magnetic confinement potential on the chip modified by rf--dressing as discussed in Lesanovsky et al.~\cite{lesanovsky:06}.  The magnetic confinement is based on an elongated Z-wire trap, the rf--field controls the splitting process. Throughout we use the same parameters as in Ref~\cite{lesanovsky:06}, and parametrize the rf-field $B_{\rm rf}=(0.5+0.3\times\lambda)$~G through the \emph{control parameter} $\lambda$ that determines the splitting distance. Note that this potential is approximated well by the function $a x^2+b x^4$, with $\lambda$-dependent parameters $a$ and $b$. Within this approach it is possible to \emph{directly} determine the experimentally accessible control field, which is often related to the tunnel coupling in a highly non-trivial fashion \cite{hofferberth:07b}.

\subsection{Condensate dynamics}

Since the number of atoms in the condensate is usually large, the many-body Hamiltonian of Eq.~\eqref{eq:ham} is approximated by expanding the field operators in the basis of \emph{orbitals}.  A famous example is the mean-field Gross-Pitaevskii (GP) equation \cite{dalfovo:99}, which is obtained by using a single orbital only. While this model describes well the dynamics of the condensate density, it does not account properly for quantum fluctuations. A widely used approximation for condensates in a double-well is provided by the generic two-mode model \cite{milburn:97,javanainen:99}, which is obtained by restricting the field operator to a left and a right orbital $\phi_L(x)$ and $\phi_R(x)$, respectively. Hence,
\begin{equation}\label{eq:fieldTM}
  \hat\Psi(x)=\hat a_L\phi_L(x)+\hat a_R\phi_R(x)\,.
\end{equation}
Within this approximation the dynamics is only related to the distribution of atoms between the left and right well, determined mainly by the tunnel coupling which is given as $\Omega=-\int dx \phi_L^*(x)\hat h(x)\phi_R(x) + h.c.$ One usually assumes that $\Omega$ can be modified by slightly varying the confinement potential of the condensate, and thus $\Omega$ allows to control the atom number dynamics. In contrast, the nonlinear intra-well energy ${\kappa}=\frac{U_0}{2}\int dx|\phi_{L,R}(x)|^4$ is assumed to be not affected by such variations. This is a reasonable assumption since $\Omega$ accounts for a tunneling process, whose efficiency depends extremely sensitive on the details of the potential, whereas $\kappa$ is expected to be rather insensitive to slight modifications of $V_\lambda$. 

For a more complete modeling of the whole splitting process, including the spatial dynamics, the ansatz for the field operator is 
\begin{equation}
  \hat\Psi(x)=\hat a_L(t)\phi_L(x,t)+\hat a_R(t)\phi_R(x,t)\,,
\end{equation}
where the orbitals depend on time now. The number distribution and orbitals are calculated self-consistently using the \emph{multi configurational time dependent Hartree for Bosons} method (MCTDHB) \cite{alon:08}. The control within this approach is now directly given by the trapping potential.

In the following, we discuss optimal condensate splitting within these two models. The simpler generic model will provide us with a more transparent picture of the underlying physics, whereas the MCTDHB approach will allow us for a realistic modeling of the splitting process.

\section{Optimization of a generic two-mode model\label{sec:Gen}}

\subsection{Model}

We start by discussing optimal squeezing within a generic two-mode model \cite{milburn:97,javanainen:99}. The Hamiltonian is obtained by plugging Eq.~\eqref{eq:fieldTM} into Eq.~\eqref{eq:ham}, and reads
\begin{equation}\label{eq:hamtwomode}
\hat H=-\frac{\Omega(t)}{2}\bigl(\hat a_L^{\dagger}\hat a_R+\hat a_L\hat a_R^{\dagger}\bigr)+\kappa\bigl(\hat a_L^{\dagger}\hat a_L^{\dagger}\hat a_L\hat a_L + \hat a_R^{\dagger}\hat a_R^{\dagger}\hat a_R\hat a_R\bigr)\,.
\end{equation}
In accordance to Ref.~\cite{milburn:97}, we have neglected all inter-well couplings, which are usually much smaller due to the small orbital overlap.  The two-mode model is then fully characterized by the tunnel coupling $\Omega(t)$. We choose units such that $2\kappa N=1$.

Eq.~\eqref{eq:hamtwomode} can be brought into a more intriguing form by introducing \emph{pseudospin operators} $\hat J_x$, $\hat J_y$, and $\hat J_z$ \cite{milburn:97}. $\hat J_x=\frac{1}{2}(\hat a_L^{\dagger} \hat a_R +\hat a_R^{\dagger}\hat a_L)$ accounts for tunneling between the left and the right well, and $\hat J_z=\frac{1}{2}(\hat a_L^{\dagger}\hat a_L -\hat a_R^{\dagger}\hat a_R)$ measures the atom number difference between the wells. We label our basis states as $|k\ra\equiv|N/2+k\ra_L|N/2-k\ra_R$, where $N$ is the total number of atoms and the subscripts denote particles in the left and right well, respectively. In this basis, the operators act on a state vector $\mathbf{C}$ whose entries are the probability amplitudes for a certain distribution of atoms between the wells. The two-mode Hamiltonian
\begin{equation}\label{eq:Hpseudo}
  \hat H=-\Omega(t)\hat J_x+2\kappa \hat J_z^2
\end{equation}
is then completely analogous to the Josephson Hamiltonian for superconductors, where $\Omega(t)$ is associated with the (time-dependent) Josephson energy and the nonlinearity $\kappa$ with the charging energy. 

For the unsplit trap we assume that all atoms reside in the bonding orbital $\phi_L+\phi_R$. Thus, the initial $\mathbf{C}$ vector corresponds to a binomial distribution for the relative atom number between left and right orbital \cite{javanainen:99},~\footnote{In this investigation, we take $\Omega(0)$ for the initial state from the self-consistent solutions of MCHB(2) calculations \cite{streltsov:06}. This will allow us for a more direct comparison with the MCTDHB results presented in Sec.~\ref{sec:Opt}. Our results, however, do not depend significantly on the exact initial state. Due to a slight quantum depletion caused by interactions we obtain a correction to the initial squeezing.}
\begin{equation}
|\mathbf{C}_0\ra=\frac{1}{2^{N/2}}\sum_{k=-N/2}^{N/2}\sqrt{\Biggl(\begin{array}{c}N\\N/2+k\end{array}\Biggr)}|k\ra\,.
\end{equation}
The standard deviation of the relative atom number between left and right well associated with this state is $\Delta n_0:=\sqrt{\langle \hat J_z^2\rangle}=\frac{\sqrt{N}}{2}$. When the trap is split, the  nonlinearity $\kappa$ becomes comparable to or larger than $\Omega$, and number fluctuations in the ground state are reduced owing to the cost of (nonlinear) energy. Finally, for the split trap we have $\kappa\gg\Omega$ and the ground state is the number eigenstate $|k=0\ra$, which can be regarded as a completely squeezed state. 

\begin{figure}
\begin{tabular}{l l}
\includegraphics[width=0.4\columnwidth]{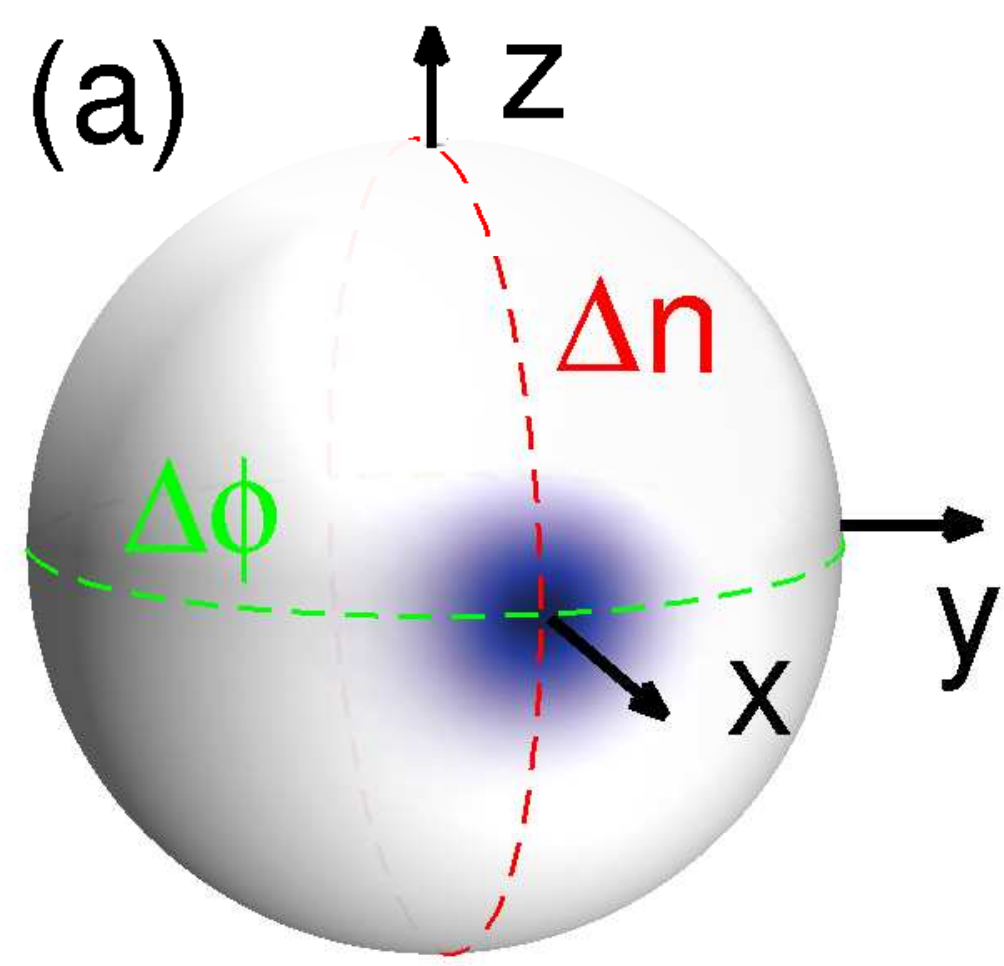}&\includegraphics[width=0.4\columnwidth]{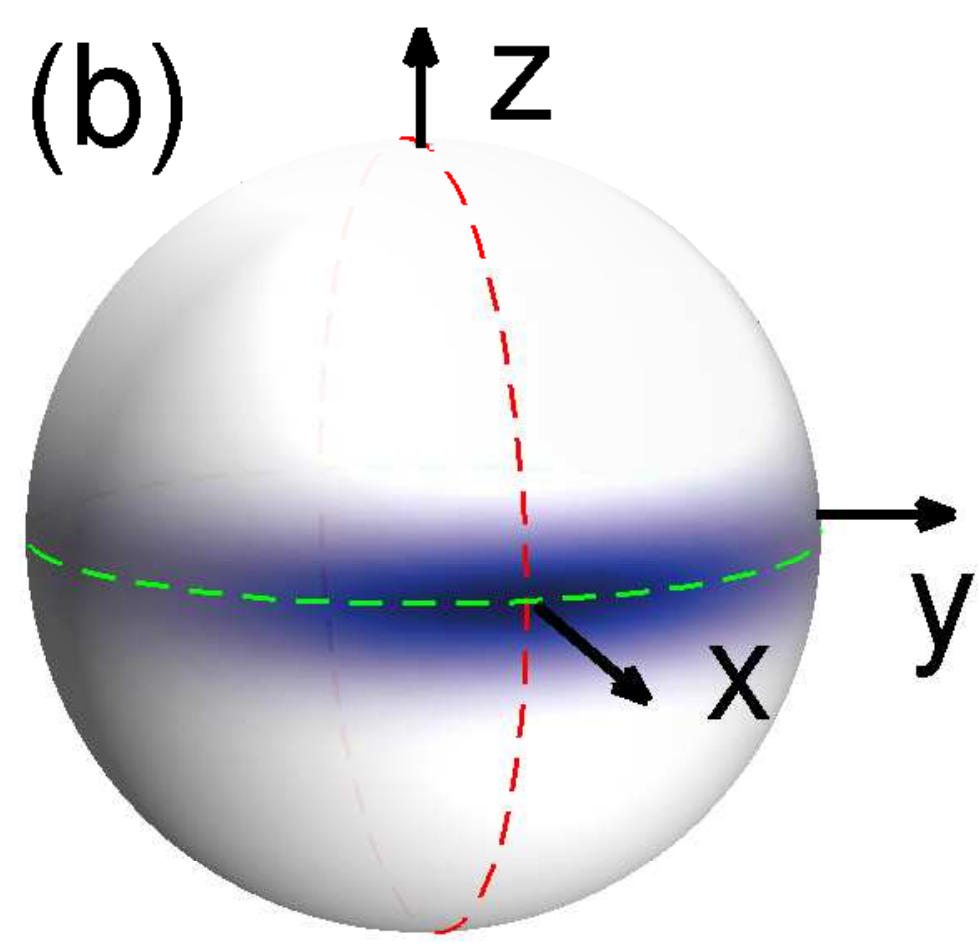}
\end{tabular}
\caption{ (Color online) Bloch sphere representation of $\mathbf{C}$. The z-axis corresponds to the number difference ($\Delta n$ and $\Delta\phi$ are number and phase fluctuations, respectively). The action of the pseudo-spin operators $\hat J_x$ and $\hat J_z$ on $\mathbf C$ corresponds to rotations about the $x$- and $z$-axis, respectively. In (a) a binomial state is shown and in (b) a number squeezed one. \label{fig:2}}
\end{figure}

A convenient representation of the atom number distribution vector $\mathbf{C}$ is on the \emph{Bloch sphere} \cite{arecchi:72}, Fig.~\ref{fig:2}. A state on the north pole would correspond to all atoms residing in the left well, and a state on the south pole to all atoms in the right well. All atoms in the bonding orbital corresponds to a state localized around $x=1$. This binomial state has equal uncertainty in number difference ($z$-axis) and in the conjugate phase observable (around the equator of the sphere). In contrast, the squeezed state shown in panel (b) has reduced number and enhanced phase fluctuations. 

\subsection{Quasi-adiabatic splitting}

\begin{figure}
 \includegraphics[width=0.8\columnwidth]{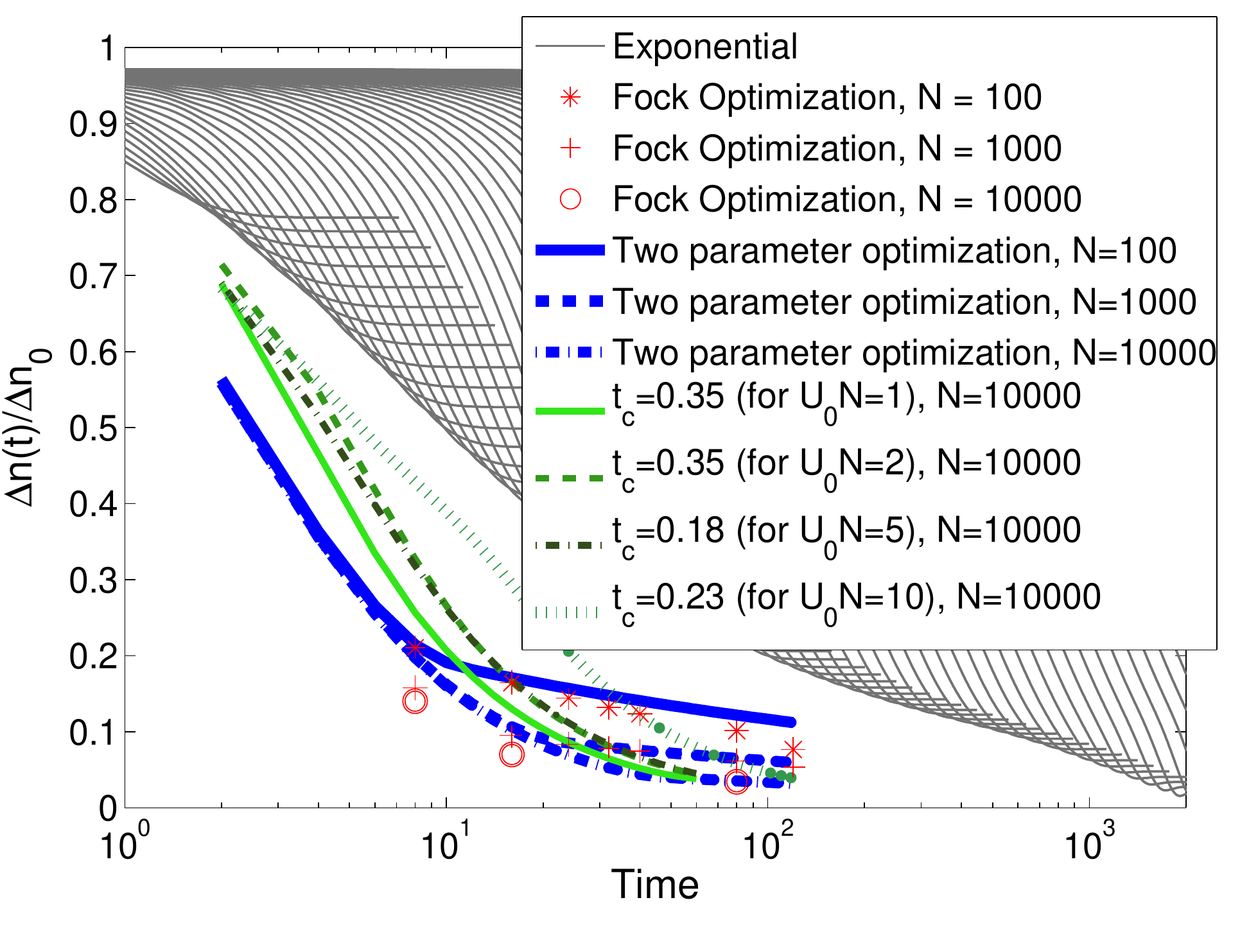}
\caption{ (Color online) Number squeezing for quasi-adiabatic exponential (gray, many lines), two-parameter (blue) and Fock optimization (symbols), for different atom numbers. For the optimized solutions we only report the final value of the number fluctuations. Both optimization strategies allow for number squeezing at least one order of magnitude faster than exponential splitting. The green lines report the limitations of the two-parameter optimization for $N=10000$ (limitations are very similar for smaller $N$). Solutions for realistic splitting within MCTDHB are only obtained for sufficiently long exponential decay constants $t_c$, reported in the panel. Smaller values of $t_c$ result in strongly oscillating condensates and reduced number squeezing. We note that in all our figures time is measured in units of $1.37$ ms and position in $\mu m$ (Rubidium units). For the generic model we use units such that $2\kappa N=1$.\label{fig:3}}
\end{figure}

Our goal is to split the condensates such that maximal number squeezing is achieved. The usual strategy to obtain squeezing is splitting slowly, such that the condensate follows adiabatically the ground state of the trap. The gray lines in Figure~\ref{fig:3} show results for a quasi-adiabatic exponential switching-off of the tunneling rate according to $\Omega(t)=\Omega(0) e^{-t/t_c}$ \cite{javanainen:99}. Depending on $t_c$, at some point adiabaticity breaks down and the fluctuations are frozen at a finite value. The inverse of the nonlinear coupling defines a minimum time $\tau_{\kappa}=1/\kappa N$ that is needed to obtain squeezing. Only for extremely long times the fluctuations approach $\Delta n=0$. 

Thus high number squeezing requires very long splitting times, and a linear improvement in squeezing is achieved only by an exponentially longer splitting time. For typical parameters, the necessary splitting time might even exceed the condensates coherence and life time.

\subsection{Two-parameter optimization}

Law et al.~\cite{law:01} suggested an alternative control scheme for achieving a much higher number squeezing in comparison to the exponential switching-off. This is achieved by applying for a given time interval a constant tunnel coupling. To understand the working principle of this control, we note that in the Bloch-sphere picture the tunnel coupling corresponds to a rotation around the $x$-axis, associated with the action of $\hat J_x$, whereas the nonlinear interaction twists the distribution around the $z$-axis, associated with the action of $\hat J_z^2$ (one-axis twisting model \cite{kitagawa:93}). The twist rate is zero at the equator, increases linearly when moving towards the poles, and the twist direction is opposite on the north and south hemisphere. For an initial binomial state, the nonlinear coupling twists the state such that it becomes a squeezed state, but along a direction which is not the equator of the Bloch sphere. In general, the number squeezing is low in this case. However, when a constant tunnel coupling is added [see first term in Eq.~\eqref{eq:Hpseudo}], it has the effect of rotating the spin squeezed state around the $x$-axis. By varying the strength and time of the tunnel coupling, an optimal number squeezing can be achieved. These optimal parameters are given by analytical formulas \cite{jin:08}.

For applying this to the splitting process, we use a control which consists of two stages:
\begin{equation}
  \Omega(t)=\Omega_0\left(1-\frac{\Omega_c}{\Omega_0}\right) e^{-t/t_c}+\Omega_c\,.
\end{equation}
First, we exponentially turn off $\Omega(t)$ with a time constant $t_c$, and in the second stage the control is kept constant. Thus, the longer $t_c$ the more the state is adiabatically squeezed in the first stage. It turned out that a more adiabatic splitting allows for more squeezing in the end, but at the price of a longer squeezing time. Optimizing for these two parameters for given time intervals, we find much better squeezing than with a pure exponential control. Results for different $N$ are displayed in Fig.~\ref{fig:3}.

\subsection{Optimal control theory \label{optsysgen}}

Optimal control theory (OCT) \cite{peirce:88,borzi.pra:02,hohenester.pra:07} provides an alternative for achieving efficient number squeezing on short time scales. OCT is a mathematical device which allows the optimization of a control objective (such as high number squeezing) for a system whose time evolution is subject to a constraint (e.g., time evolution governed by the two-mode model). In general, the ``optimal control'' is obtained numerically through an iterative procedure \cite{peirce:88,borzi.pra:02}. We start by defining a \emph{cost function} which quantifies the optimization target. For the case of number squeezing it reads\footnote{In $\hat H$ we additionally make the replacement $\Omega\rightarrow\Omega_s^2$ in order to ensure positive tunnel couplings.}
\begin{equation}\label{eq:costGen}
J(\mathbf{C},\Omega_s)=\la  \mathbf{C}(T)| \hat J_z^2|\mathbf C(T)\ra+\frac{\gamma}{2}\int_0^T (\dot{\Omega}_s)^2 dt\,,
\end{equation}
where $T$ is the terminal time of the control sequence. The second term, which is weighted by the regularization parameter $\gamma$, penalizes rapidly varying tunnel couplings and is needed to make the control problem well posed. We choose the value $\gamma=10^{-3}$ throughout this work, such that the first term dominates. $J(\mathbf{C},\Omega_s)$ is now optimized under the constraint that $\mathbf{C}$ obeys the time evolution of the Schr\"odinger equation, with the Hamiltonian given in Eq. \eqref{eq:Hpseudo}. This is achieved by introducing a Lagrange multiplier $\tilde{\mathbf{C}}$ for the constraints. We define a Lagrange function
\begin{equation}\label{eq:Lag}
L(\mathbf{C},\tilde{\mathbf{C}},\Omega_s)=J(\mathbf{C},\Omega_s)+\mbox{Re}\Bigl<\tilde{\mathbf{C}}, i\dot{\mathbf{C}}-\hat H \mathbf{C}\Bigr>\,,
\end{equation}
where $\tilde{\mathbf{C}}$ is also called the \emph{adjoint} variable. The bracket stands for $\la \mathbf u,\mathbf v\ra=\int_0^T \mathbf \langle\mathbf u(t)|\mathbf v(t)\rangle\,dt$. $L(\mathbf{C},\tilde{\mathbf{ C}},\Omega_s)$ has a saddle point at the minimum of $J(\mathbf{C},\Omega_s)$. Thus, the Fr\'echet derivatives with respect to $\mathbf{C}$, $\tilde{\mathbf{C}}$ and $\Omega_s$ must all be zero for the optimal control. From this we obtain Euler-Lagrange equations, together with the initial and terminal conditions,
\begin{subequations}
\begin{eqnarray}\label{eq:controlsysa}
i|\dot{\mathbf{C}}\rangle&=&\hat{H}|\mathbf{C}\rangle\,,\quad\mathbf{C}(0)=\mathbf{C_0}\\\label{eq:controlsysb}
i|\dot{\tilde{\mathbf{C}}}\rangle&=&\hat{H}|\tilde{\mathbf{C}}\rangle\,,\quad
i|\tilde{\mathbf{C}}(T)\rangle=2\hat J_z^2 |\mathbf C(T)\rangle\\
\gamma \ddot{\Omega}_s&=&2\Omega_s \mbox{Re}\langle \tilde{\mathbf{C}}|\hat J_x |\mathbf C \rangle\,.
\end{eqnarray}
\end{subequations}
The equation in the third line is subject to the (desired) boundary conditions $\Omega_S(0)=\Omega_S^0$ and $\Omega_S(T)=0$ and it is fulfilled for the optimal control. For any other, non-optimal control the third equation does not hold. Instead, the so-called \emph{control equation}
\be\label{eq:gradient}
\nabla J=-\gamma \ddot{\Omega}_s+2\Omega_s \mbox{Re}\langle \tilde{\mathbf{C}}|\hat J_x |\mathbf C \rangle\,.
\ee
applies, which can be interpreted as the gradient of the cost functional with respect to a variation of the control field. It allows for the implementation of an iterative numerical scheme to determine the optimal control.

One can thus use the following algorithm to iteratively improve the control: For an initial guess, one first calculates the forward equation Eq.~\eqref{eq:controlsysa}, and thereafter the backward equation Eq.~\eqref{eq:controlsysb}. From the gradient in Eq.~\eqref{eq:gradient}, a new search direction is found, from which an optimization scheme, such as the nonlinear conjugate gradient or quasi-Newton one \cite{borzi.pra:02,bertsekas:99}, determines a new control. For the time propagation we use a Crank-Nicolson scheme similar to Ref.~\cite{javanainen:99}.
 
The direct implementation of the above optimization scheme can lead to problems. One of them is the appearance of discontinuities of $\Omega(t)$ at the initial and final time. Another one is that the algorithm is sometimes unable to optimize the global structure of the control but sticks very close to the initial guess. Both problems can be circumvented by using a more refined approach \cite{winckel:08}: instead of the usual $L^2$-formulation of the control problem, we employ a control $\lambda\epsilon H^1$, i.e., in the space of $L^2$ functions with weak derivative again in $L^2$. The Lagrange function (and all inner products appearing in the optimization algorithm) has then to be reformulated using the $H^1$ inner product $(u,v)_{H_1}=(\dot u,\dot v)_{L^2}$. In particular, we get  $\frac{\gamma}{2} (\lambda,\lambda)_{H^1}$ for the regularization term. While the forward and adjoint equations remain unchanged within this approach, the gradient is given now by a Poisson equation
\be\label{eq:poisson}
-\frac{d^2}{dt^2}\nabla J=-\gamma \ddot{\Omega}_s-2\Omega_s \mbox{Re}\langle \tilde{\mathbf{C}}|\hat J_x |\mathbf C \rangle\,,
\ee
which distributes local changes of the rhs to the complete time interval. Since second time derivatives of $\Omega_s$ appear on both sides of the equation, the gradient has the same degree of smoothness as the control. Further, the final and terminal conditions of the control are fulfilled exactly. In the context of the GP-equation it was demonstrated that this $H^1$ approach leads to a smoother control and quicker convergence than the common $L^2$ one \cite{winckel:08}.

While the standard $L^2$ approach is sufficient for our optimizations within the generic model, we will find clear-cut advantages of the $H^1$ formulation in case of the realistic simulations.

\begin{table}
 \caption{ (Color online) Definition of Josephson and Fock regime used in the text. There exists also the Rabi regime ($\Omega/\kappa\gg N$), which, however, does not play a role in the context of number squeezing. \label{tab:regimes}}
 \begin{ruledtabular}
\begin{tabular}{lll}

Regime & Criterion& Characteristics \\
\hline
Josephson&$1/N\ll\Omega/\kappa\ll N$& Competition between\\
&& $\Omega$ and $\kappa$\\
Fock&$\Omega/\kappa\ll1/N$& Non-linear interaction \\ && $\kappa$ dominates
\end{tabular}
 \end{ruledtabular}
 \end{table}

\subsection{OCT Results}

In a bosonic Josephson junction one distinguishes between the Rabi-, Fock- and Josephson regimes \cite{leggett:01,pezze:05}, which are characterized by different values of $\Omega/\kappa$ listed  in Table~\ref{tab:regimes}. In the Josephson regime there is a competition between tunneling and the non-linear interaction $\kappa$, while in the Fock regime $\kappa$ dominates.
It turns out that our OCT fields decisively depend on the initial guess for the tunnel coupling $\Omega(t)$, although the final states have similar number squeezing and are both suited for atom interferometry purposes, as will be discussed in section~\ref{sec:AI}. We essentially find two types of solutions, which can be classified according to physical regimes discussed above as \emph{Josephson} and \emph{Fock} solutions.

\begin{figure}
\begin{tabular}{l}
(a)\\
\includegraphics[width=0.8\columnwidth]{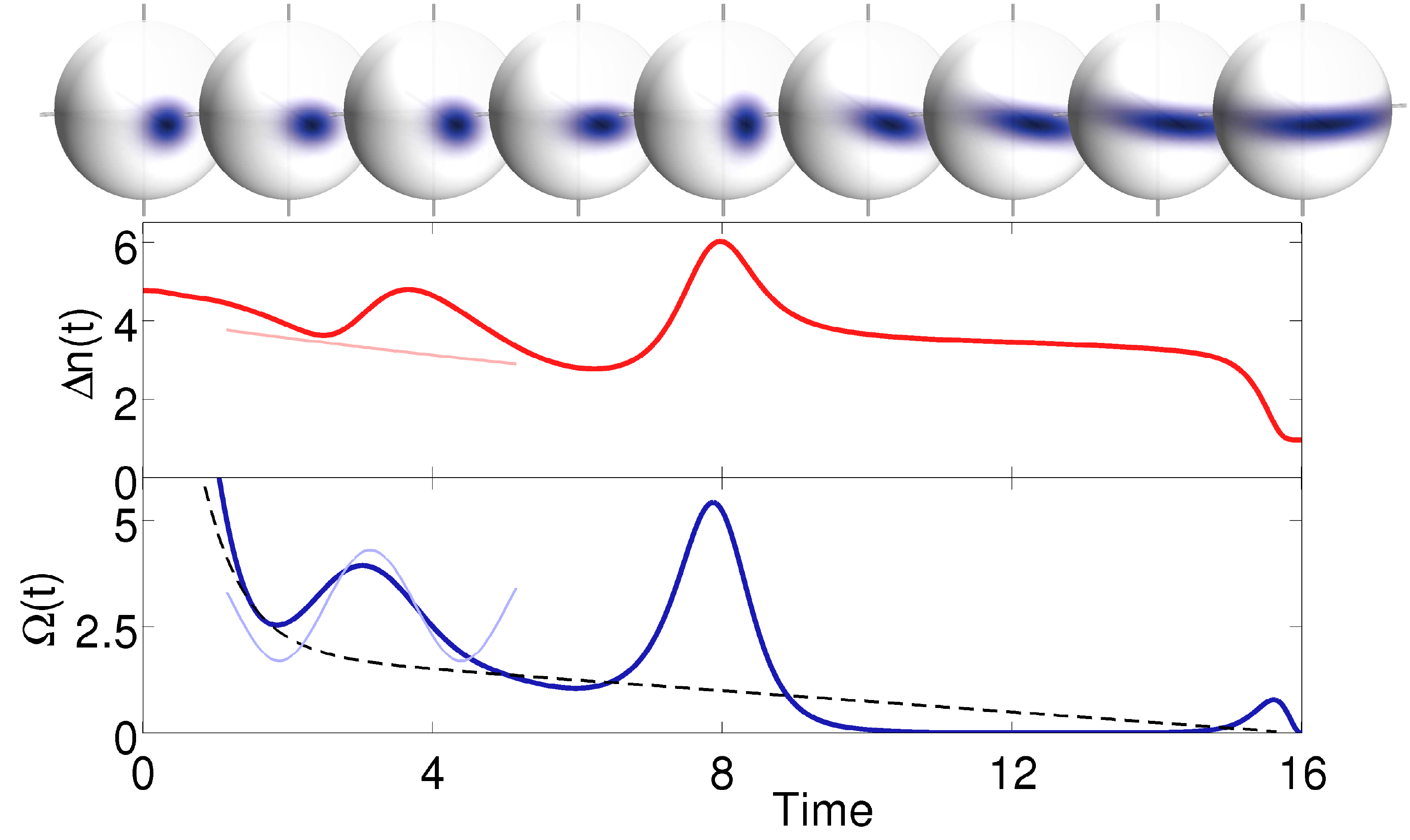}\\
(b)\\
\hspace{0.1cm}\includegraphics[width=0.74\columnwidth]{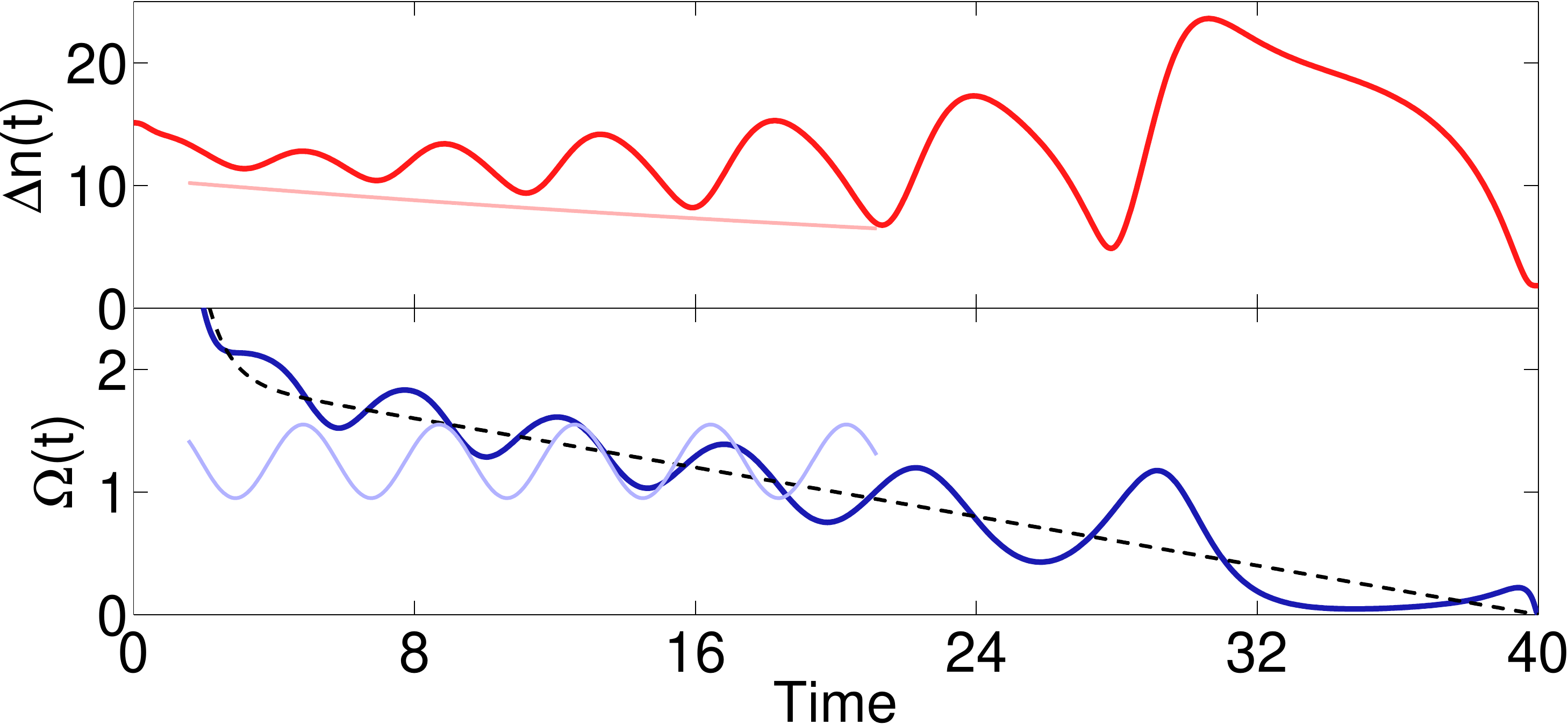}
\end{tabular}
\caption{ (Color online) Typical results from \emph{Josephson} OCT for the generic two-mode model for different splitting times and (a) $N=100$, (b) $N=1000$. The initial (optimized) $\Omega$ is shown by the dashed (solid) line in the lower panels of the figures. The upper panels of the figures show the number fluctuations. The bright lines show estimates from a parametric oscillator model, see Sec.~\ref{subsec:interpret}. (a) The Bloch spheres demonstrate the state for different times. \label{fig:4}}
\end{figure}

For the \emph{Josephson solution}, the initial guess (dashed lines in Fig.~\ref{fig:4}) is chosen such that after a first exponential decrease of $\Omega$ the control decreases linearly. In this second stage, the system is in the Josephson regime with a competition between tunnel coupling and nonlinear interaction.
With this initial guess, OCT comes up with an optimized $\Omega$ that shows oscillations in the Josephson regime, as shown in Fig.~\ref{fig:4}. Also the fluctuations $\Delta n$ oscillate, and at the final time the system is brought to a highly squeezed state. 

\begin{figure}
\begin{tabular}{l}
(a)\\
\includegraphics[width=0.8\columnwidth]{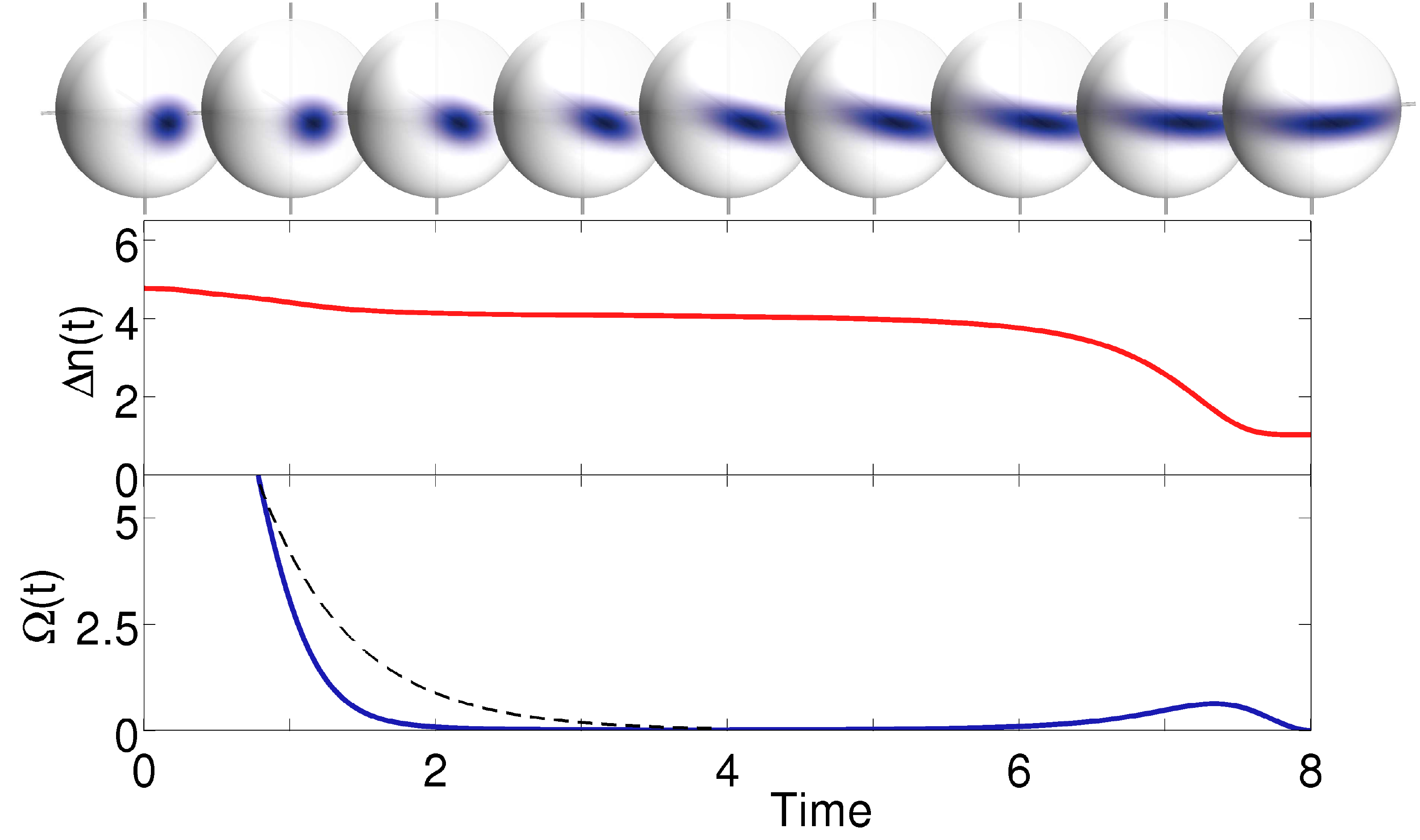}\\
(b)\\
\hspace{0.1cm}\includegraphics[width=0.74\columnwidth]{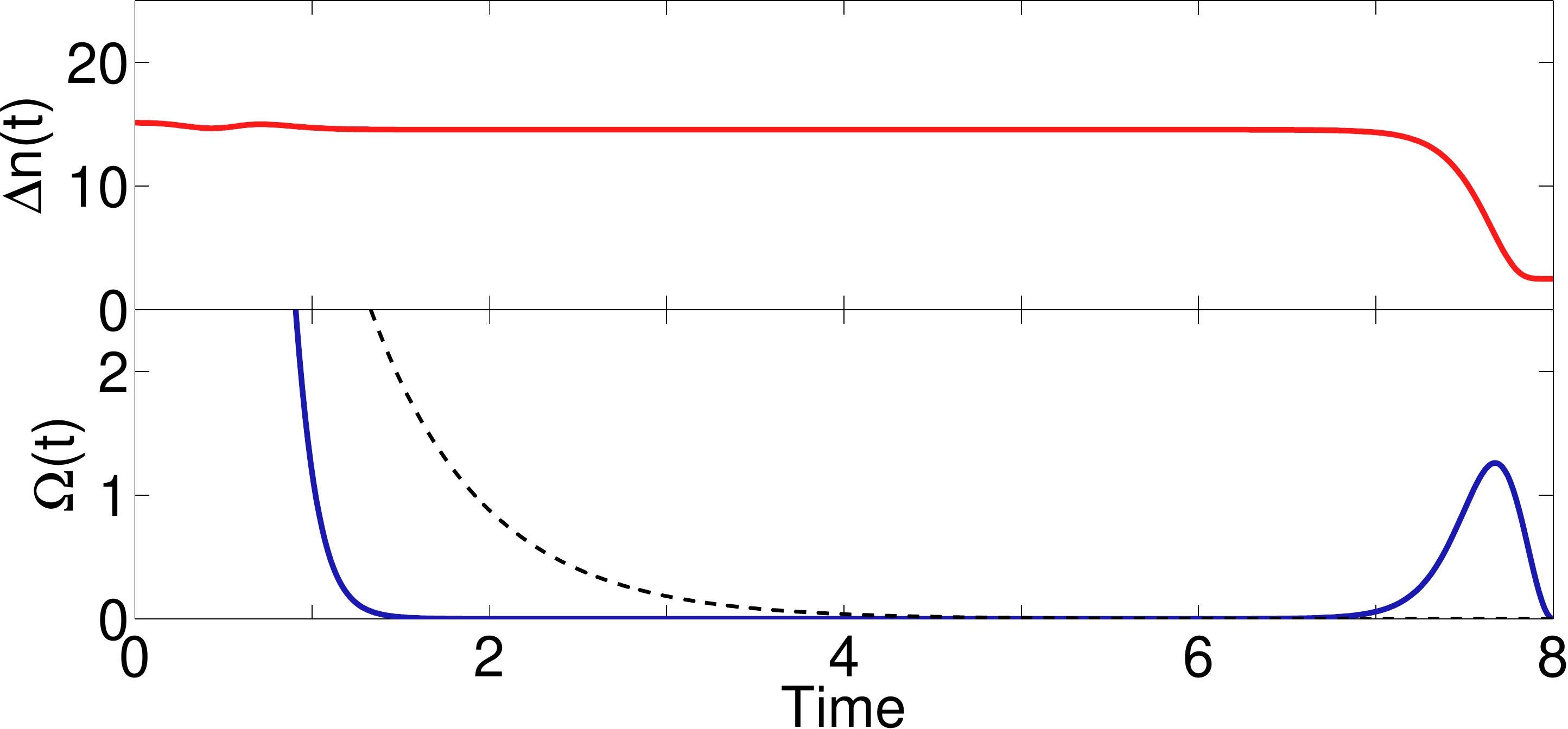}
\end{tabular}
\caption{ (Color online) Typical results from \emph{Fock} OCT for the generic two-mode model for (a) $N=100$, (b) $N=1000$. The initial (optimized) $\Omega$ is shown by the dashed (solid) line in the lower panels of the figures. The upper panels of the figures show the number fluctuations. (a) The Bloch spheres demonstrate the state for different times.\label{fig:5}}
\end{figure}

For the \emph{Fock} solution the initial guess is a simple exponential decay (dashed lines in Fig.~\ref{fig:5}). Here, the optimized $\Omega$, displayed in Fig.~\ref{fig:5}, is most of the time very small (and thus can be classified to be of Fock type). Only for two short time periods it rises to larger values. Comparison between exponential, Josephson, and Fock OCT splitting is made in Fig.~\ref{fig:12}. Best squeezing is achieved within the last approach. 

\subsection{Interpretation\label{subsec:interpret}}

\emph{Josephson optimization.}---We next interpret the oscillating behavior in the Josephson control of Fig.~\ref{fig:4}. To this end, we exploit that for large atom numbers $N$ the state vector $\mathbf C$ can be approximately treated as a continuous variable \cite{javanainen:99}. We consider $C(k)$ as the atom-number wave function and $k$ as the number difference between the left and right well. The matrix elements of Eq.~\eqref{eq:hamtwomode} are Taylor expanded in powers of $\frac 1N$, and $C(k\pm 1)$ is Taylor expanded around $k$. 
We then obtain a Schr\"odinger equation for the variable $C(k)$ 
\begin{equation}\label{eq:twomodeosc}
  i\dot C(k)=\left[-\frac{\Omega(t)N}{4}\frac{\partial^2}{\partial k^2}+\left(\frac{\Omega(t)}{N}+2\kappa\right)k^2\right]C(k)\,,
\end{equation}
which can be identified as a parametric oscillator in the variable $k$
\cite{messiah:65}. One could think of the parametric oscillator as a harmonic oscillator with constant frequency but with a driving term proportional to $k^2$. Thus, parametric amplification translates to a nonlinear driving process, where the fluctuations (and not the mean atom number difference) are subject to parametric amplification.

In appendix \ref{app:A} it is shown that if the oscillator is driven by an oscillating tunnel coupling $\Omega(t)=\Omega_0+\Omega_1\cos{(\omega t)}$ (we assume $\Omega_1\ll\Omega_0$) with approximately twice the resonance frequency $\omega_{res}=\kappa N+\Omega_0$, the envelope for the width of the initial number distribution $C(k)$ decays exponentially.  For the envelope of the number fluctuations we obtain \cite{scully:97}
\be\label{eq:env}
(\Delta n)_{\rm envelope}\sim\Delta n_0 \exp\left[{-\frac{\kappa N\Omega_1}{2(\Omega_0+\kappa N)} t}\right]\,.
\ee

The bright lines in Fig.~\ref{fig:4} show results of this model: the oscillation period and shape of the control indeed mimic $\Omega_{\rm OCT}(t)$, and the envelope of the number fluctuations is in good qualitative agreement with $\Delta n_{\rm OCT}(t)$. The fluctuations are reduced to values of approximately $\Delta n/\Delta n_0\sim0.3-0.5$, depending on the number of atoms. 
During the final stage of the control, nonlinear coupling dominates over tunneling, and the approximations introduced above fail. Squeezing in this interaction-dominated regime can be understood in terms of the one-axis twisting model discussed above.

\emph{Fock optimization.}---The physics underlying Fock optimization can be regarded as a generalization of the model underlying the two-parameter optimization \cite{law:01}. 
The results from Fig.~\ref{fig:5} can then be interpreted as follows. After the exponential splitting, we end up in a squeezed state which, however, is not a number-squeezed state. The first $\Omega$ pulse after the exponential decay rotates the squeezed state towards the equator, but a little tilt remains, which leads in the ensuing time evolution to squeezing caused by the non-linear interactions. During this stage $\Omega$ is almost zero. In the end, a second $\Omega$ pulse rotates the state to the equator, such that the number squeezing becomes maximized. 

In Fig.~\ref{fig:3} we compare the two-parameter optimization with Fock OCT. In a wide regime of splitting times and atom numbers, the squeezing achieved within both schemes agrees, except for small atom numbers and long splitting times, and large atom numbers and short splitting times. This is due to the fact that for the constant control scheme \cite{law:01} there exists an optimal squeezing and an optimal squeezing time, which are both directly proportional to $N$ \cite{jin:08}. In case of OCT however, we are not restricted to a single control strategy, such as the mere $\Omega_c$ value. As a consequence, OCT can come up with more improved control strategies and performs better than the more simple schemes.

\section{Optimization of a realistic model \label{sec:Opt}}

We next proceed to a more realistic model for the splitting process, which captures also the spatial dynamics and relies on the control parameter used in experiments, related to the double well separation of the trapping potential. As consequence, we can also control the oscillations of the condensate when it is split fast. Methods for including the orbitals describing the spatial dynamics rely on variational principles, as for example in Ref.~\cite{menotti:01} where a Gaussian ansatz for the orbitals was used. 

In our approach we employ the \emph{multi configurational time dependent Hartree for Bosons} method (MCTDHB) \cite{alon:08,streltsov:07}, where the orbitals are included self-consistently, without any assumptions about their shape. Although the MCTDHB framework is formulated for an arbitrary and fixed number of orbitals, we will stay with two, as they are sufficient to account for condensate splitting.

\subsection{ Doublewell potential \label{sec:Dou}}

In our calculations we consider condensates in elongated magnetic traps with a tight transverse and a weak longitudinal confinement as can be manipulated easily on atom chips. For the transverse confinement, we use the 2-d potential given in Ref.~\cite{lesanovsky:06}. In the longitudinal direction we take an harmonic confinement with trap frequency $\omega_{\parallel}$. Condensate splitting is achieved by modifying the trap potential along a transverse direction, through the confinement potential discussed in Sec. II B and Ref.~\cite{lesanovsky:06}. The trap is then split along the $x$-direction by varying the radio-frequency field $B_{\rm rf}(\lambda)$, where $\lambda=-\frac 23$ corresponds to the unsplit trap which can be approximated by an harmonic well with a transverse trap frequency $\omega_\perp=2\pi\cdot2$ kHz.

In our model we assume that the splitting direction is decoupled from the other ones. For typical aspect ratios $\omega_{\perp}/\omega_{\parallel}$, the dynamics in the weakly confined longitudinal direction is much slower than in the splitting direction. The 3D interaction strength is given in a contact potential approximation $U_0^{3D}=4\pi\hbar^2 a/m$, where $a$ is the  s-wave scattering length and $m$  the mass of an atom. For calculating the 1D interaction parameter $U_0$, we consider the stationary GP equation \cite{dalfovo:99} in 3D with the confinement potential $v(\mathbf r)=v_x(x)+v_y(y)+v_z(z)$ and define the single particle Hamiltonians $\hat{h}_i=-\frac 12\nabla_i^2+v_i$ ($i=x,y,z$). We assume that the wave function factorizes, $\phi(\mathbf r)=\phi_x(x)\phi_y(y)\phi_z(z)$, and we integrate out the $y$ and $z$ coordinates. $\phi_x$ then satisfies a non-linear Schr\"odinger equation with the Hamiltonian
\be
H_x=h_x\phi_x+U_0^{3D} (N-1)W_yW_z|\phi_x|^2\phi_x\,,
\ee
where $W_y=\int dy |\phi_y(y)|^4$, with an analogous expression for $W_z$. $\phi_y$ and $\phi_z$ satisfy similar equations with $H_y$ and $H_z$, respectively. Employing imaginary time-propagation simultaneously for $\phi_x$, $\phi_y$ and $\phi_z$, we obtain the ground state of the trap. The 1D interaction parameter for splitting in $x$-direction is then given as $U_0 = U_0^{3D}W_yW_z$.

\begin{figure}
\begin{tabular}{l}
\includegraphics[width=0.8\columnwidth]{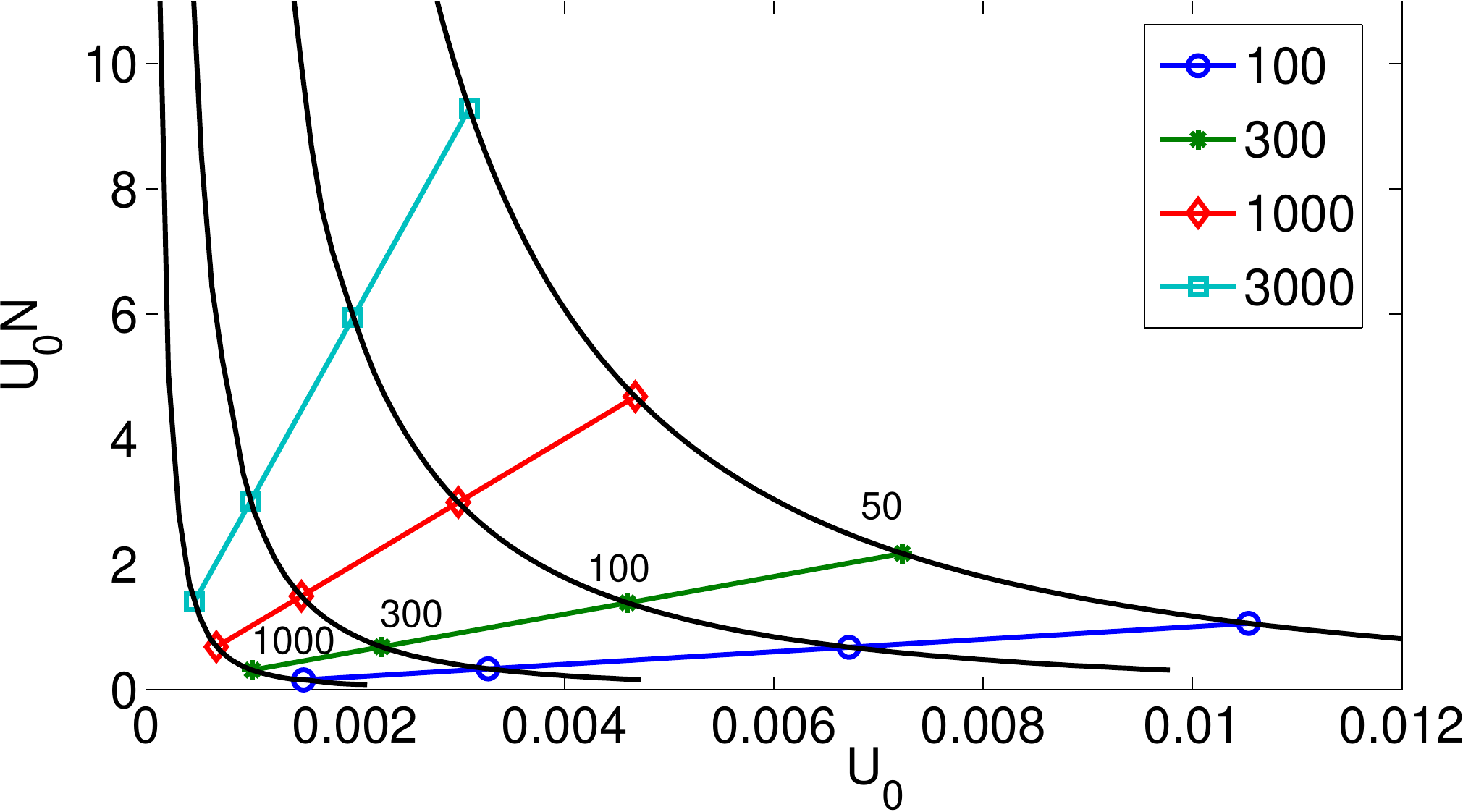}
\end{tabular}
\caption{ (Color online)  Inverse adiabatic squeezing time scale $U_0 N$ versus 1D-interaction strength $U_0$ for several $N$, where each point along a straight line corresponds to a different aspect ratio (connected by black lines to points with same aspect ratio), as given in the figure. The straight lines are guides to the eye. \label{fig:6}}
\end{figure}

In Fig.~\ref{fig:6} we plot $U_0 N$, which determines the squeezing time scale for the quasi-adiabatic exponential splitting, versus the 1-d interaction parameter $U_0$. The straight lines correspond to a fixed atom number $N$ taken for various values of the aspect ratio of the trap $\omega_{\perp}/\omega_{\parallel}=50,100,300,1000$. For small aspect ratios $U_0$ is largest. It is apparent from the figure that neither $U_0$ nor $U_0 N$ do increase linearly with the atom number $N$. Instead, for higher $N$ the wave function spreads out more along the longitudinal direction, and thus $U_0$ is reduced. Therefore, the value $U_0 N=1$ as assumed in most of our optimizations is quite realistic. The chemical potential is between $\mu\sim2\pi\cdot2$ kHz (large aspect ratio) and $\mu\sim2\pi\cdot4$ kHz (small aspect ratio).

\subsection{MCTDHB}

\subsubsection{Equations}

The MCTDHB equations are obtained by making an ansatz for the many-body wave function in a basis of two orbitals, which are allowed to depend on time, and apply variational calculus. In the following we summarize the MCTDHB equations \cite{alon:08}.  Since our trapping potential is symmetric, the orbitals exhibit \emph{gerade} ($\phi_g$) and \emph{ungerade} ($\phi_e$) symmetry.

The orbitals equations read
\bea\label{eq:MCTDHBorbs}
i\dot{\phi}_g&=&\hat{\mathbf{P}}\Bigl[\hat{h}\phi_g+(f_g^g|\phi_g|^2+f_g^e|\phi_e|^2)\phi_g+\tilde f_g\phi_g^*\phi_e^2\Bigr]\,,\quad\\		
i\dot{\phi}_e&=&\hat{\mathbf{P}}\Bigl[\hat{h}\phi_e+(f_e^g|\phi_g|^2+f_e^e|\phi_e|^2)\phi_e+\tilde f_e\phi_e^*\phi_g^2\Bigr]\,.\quad\nonumber
\eea
 The coefficients are given by
\bea \label{eq:coeffs}
&&f_k^k=U_0\{\mathbf{\rho}\}_{kk}^{-1}\rho_{kkkk}\,,\quad f_k^q=2U_0\{\mathbf{\rho}\}_{kk}^{-1}\rho_{kqkq}\,,\non
&&\tilde f_k=U_0\{\mathbf{\rho}\}_{kk}^{-1}\rho_{kkqq}\,,
\eea
where $k$ is either $g$ or $e$, and $q$ the opposite. The one- and two-body reduced densities \cite{sakmann:08} $\rho_{kq}$ and $\rho_{kqlm}$, respectively, are given by 
\bea
&&\rho_{kk}=\langle\mathbf{C}|\hat a_k^{\dagger}\hat a_k|\mathbf C\rangle\,,\quad \rho_{kkkk}=\langle\mathbf{C}|\hat a_k^{\dagger}\hat a_k\hat a_k^{\dagger}\hat a_k|\mathbf C\rangle\,,\\
&&\rho_{kqkq}=\langle\mathbf{C}|\hat a_k^{\dagger}\hat a_k\hat a_q^{\dagger}\hat a_q|\mathbf C\rangle\,,\quad\rho_{kkqq}=\langle\mathbf{C}|\hat a_k^{\dagger}\hat a_k^{\dagger}\hat a_q\hat a_q|\mathbf C\rangle\,.\nonumber
\eea
Note that because the orbitals have different parity, the one-body reduced density is diagonal and all matrix elements of the two-body reduced densities with uneven combinations of indices vanish \cite{streltsov:06}.   We distinguish between two types of nonlinearities: the terms of type $f_k^k$ and $f_k^q$ are GP-like, where the orbitals experience a real potential proportional to the moduli of the orbitals. The terms of type $\tilde f_k$ are complex contributions, and couple the two orbitals in a more complicated fashion. $\hat{\mathbf{P}}= 1-|\phi_g\ra\la\phi_g|- |\phi_e\ra\la\phi_e|$ is a projector which guarantees that the orbitals stay normalized throughout. 

The number distribution vector obeys the equation
\be\label{eq:TMMC}
i\frac{\partial \mathbf{C}(t)}{\partial t}=\mathcal{H}\mathbf{C}\,.
\ee
The two-mode Hamiltonian is given as
\be\label{eq:TMMCham}
\mathcal{H}=-\Omega\hat{J}_x+\frac{1}{2}\sideset{}{'}\sum_{k,q,l,m}\hat a_k^{\dagger}\hat a_q^{\dagger}\hat a_l\hat a_m W_{kqlm}\,,
\ee
where the prime indicates that the sum only runs over even combinations of indices, which is due to the parity of the orbitals. We have introduced annihilation and creation operators $\hat a_k^{(\dagger)}$ for the gerade and ungerade orbitals. Further, we have $\hat{J}_x=\frac 1 2 (\hat a_g^{\dagger}\hat a_g -\hat a_e^{\dagger}\hat a_e)$ and $\Omega=\la \phi_e|\hat{h}|\phi_e\ra-\la \phi_g| \hat{h}|\phi_g\ra$. The integrals
\be\label{eq:mvs}
W_{kqlm}=U_0\int dx \phi_k^*(x)\phi_q^*(x)\phi_l(x)\phi_m(x)\,,
\ee
are the two-body matrix elements. 
Eqs.~\eqref{eq:MCTDHBorbs} and \eqref{eq:TMMC} are now coupled equations which have to be solved simultaneously.

\subsubsection{Ground state properties \label{subsubsec:ground}}
  			
We obtain the ground state of the trap for a fixed splitting $\lambda$ by applying imaginary time-propagation to Eq.~\eqref{eq:MCTDHBorbs} and simultaneously diagonalising Eq.~\eqref{eq:TMMC}. This gives us the MCHB groundstates.

\begin{figure}
\centering
\includegraphics[width=0.6\columnwidth]{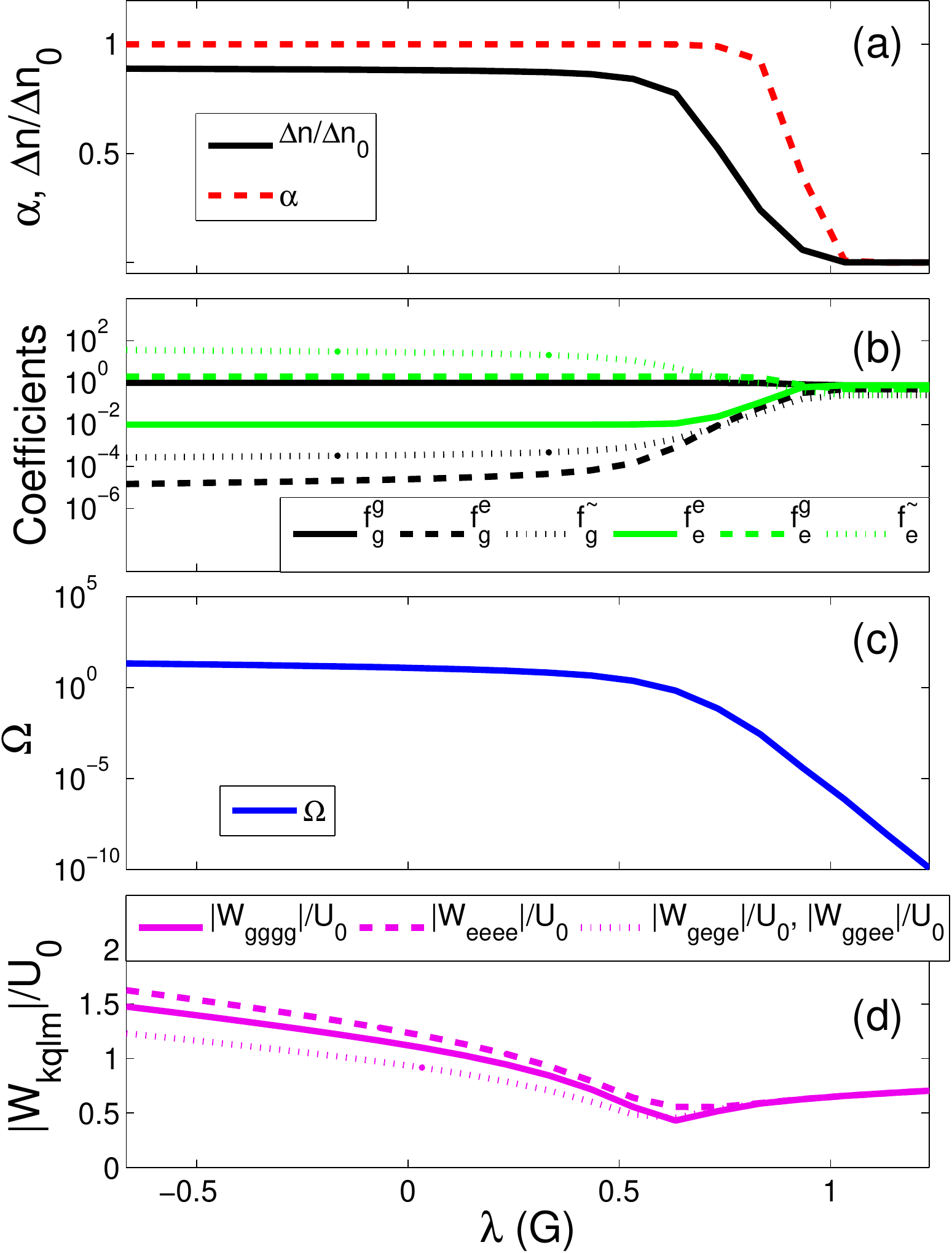}
\caption{ (Color online) (a) Mean population difference  $\alpha=(\rho_{gg}-\rho_{ee})/N$ of the orbitals normalized by $N$ ($\alpha$ is the coherence factor) and number fluctuations , (b) nonlinear coefficients of the orbitals equations from Eq.~\eqref{eq:coeffs}, (c) the tunnel coupling, and (d) the two-body matrix elements of Eq.~\eqref{eq:mvs}. Parameters are $U_0 N=1$ and $N=100$ (in the following figures we always use these values, unless stated otherwise).\label{fig:7} }
\end{figure}

Consider first the unsplit trap: The gerade orbital is populated to more than $99\%$ [Fig.~\ref{fig:7} (a)]. The only significant nonlinear coupling in Eq.~\eqref{eq:MCTDHBorbs} is due to $f_g^g\approx U_0 N$ [Fig.~\ref{fig:7} (b)]. This reflects that the ground state of a harmonic trap is well described by the GP equation, and the quantum depletion out of the condensate is low \cite{streltsov:06}. At the same time, the ungerade orbital is affected strongly, through the coefficient $\tilde f_e$, such that the energy gap between the orbitals becomes larger in comparison to the harmonic oscillator states. For increasing interaction strength the gap decreases, and the excitations to the ungerade orbital are enhanced. We note that the structure of the term proportional to $\tilde f_e$ is reminiscent of the Bogoliubov-de Gennes equations, which describe weak excitations out of the condensate \cite{leggett:01}. 

When the potential is split into two separate wells, the energy splitting decreases towards zero, and so does the tunnel coupling [Fig.~\ref{fig:7} (c)]. In addition, the ungerade orbital becomes more and more populated, as shown by the red dashed line in Fig.~\ref{fig:7} (a).  For the split trap, the wave function $\phi_{g,e}=(\phi_L\pm\phi_R)/\sqrt{2}$ can be decomposed into \emph{left} and \emph{right} orbitals, which are solutions of Gross-Pitaevskii equations with the nonlinearity for $\sim N/2$ atoms. In Fig.~\ref{fig:7} (d) we plot the two-body matrix elements from Eq.~\eqref{eq:mvs}. They drop by a value of about two during splitting, and are all equal only for the split trap.\footnote{In our comparison of the generic two-mode model and MCHB, we neglect the effect of different $W_{kqlm}$ before and after splitting, which would lead to a slight renormalization of the tunnel coupling \cite{javanainen:99,ananikian:06}. We use the value of $W_{gggg}$ of the split trap, which deviates by a factor of approximately two from the value for the unsplit trap. Considering that the two-body matrix elements of left-right and gerade-ungerade orbitals differ by a factor of two, we are left with $\kappa=U_0/2$.}

 Number fluctuations in the split condensate are shown in Fig.~\ref{fig:7} (a). The observable for detecting an atom on the left side (i.e., in the left well of the split trap) reads
\be
\hat{N}_L=\int_{-\infty}^{0} dx \hat{\Psi}^{\dagger}\hat{\Psi}=N/2 + d \ah_g^{\dagger}\ah_e+d^*\ah_e^{\dagger}\ah_g\,,
\ee
where $d=\int_{-\infty}^{0}\phi_g^*(x)\phi_e(x)\,dx$ is the overlap integral over the negative half axis. A similar expression is obtained for $\hat N_R$ by restricting the integration to the positive half axis. The number difference operator is then given by $\Delta\hat{N}/2=(\hat{N}_L-\hat{N}_R)/2=d \ah_g^{\dagger}\ah_e+d^*\ah_e^{\dagger}\ah_g$, where we have exploited the symmetry of the orbitals. For the split trap, where $\phi_{g,e}=(\phi_L\pm\phi_R)/\sqrt{2}$ holds, we obtain $d=\frac 12$ and $\Delta \hat N/2$ corresponds to $\hat J_z=(\hat a_L^{\dagger}\hat a_L-\hat a_R^{\dagger} \hat a_R)/2$, otherwise $d<\frac 12$. For a dynamic splitting of the trap, $d$ can acquire an additional phase, which, however, is irrelevant when computing any expectation value. 

\subsection{Optimality system}
   
The cost functional consists of a squeezing and a regularization term, analogous to the cost function for the generic model in Eq.~\eqref{eq:costGen}. Additionally we require that the spatial dynamics is stationary at the final time to avoid oscillations of the condensates in the split trap \cite{hohenester.pra:07}. It turns out that this can be achieved by trapping the orbitals in the GP ground and first excited state, respectively, which become the ``desired'' orbitals $\phi_g^d$ and $\phi_e^d$ of our optimization approach. We find that, at least for moderate interactions, these orbitals are stationary, independent on the precise value of the number fluctuations. We then have	
\bea\label{eq:costMCTDHB}
&&J(\phi_g,\phi_e,\mathbf{C},\lambda) = \frac{\gamma_1}{2} \langle\mathbf{C}(T)|\frac{\Delta\hat{N}^2}{4}|\mathbf{C}(T)\rangle   \non
&&+ \frac{\gamma_2}{2}[2-|\la \phi_g(T)|\phi_g^d\ra|^2-|\la \phi_e(T)| \phi_e^d\ra|^2]\non  
&&+ \frac{\gamma}{2}\int_0^T\bigl[\dot{\lambda}(t)\bigr]^2 dt\,,
\eea
where  $\gamma_1$, $\gamma_2$ and $\gamma$ weight the different control objectives. For the control $\lambda$ we choose fixed initial and terminal conditions, i.e., we require an initial unsplit trap and a final split trap. 
The optimality system for MCTDHB, which is quite involved, is derived analogous to Sec.~\ref{optsysgen}. Details are given in Appendix~\ref{app:B}.

\subsection{Numerical implementation}

The MCTDHB equations are highly nonlinear equations, due to the orbitals nonlinearities, the state vectors dependence on tunnel coupling and two-body matrix elements, and the projectors. For the time integration we use a Modified Crank-Nicolson method \cite{Thomee:97} with Newton iterations at each time step. Details are given in appendix \ref{app:C}. Using an implicit scheme is very advantageous in connection with OCT, since preservation of the norm is crucial. This is achieved here with finite step size and accuracy $\mathcal{O}(\Delta t^3)$. For the backward equations, we can use the wave function from forward integration due to the constant time step. 

A very useful criterion for testing the correct implementation of the optimality system, which is usually quite involved, is as follows. We choose test functions $u$ and $v$, and compare the finite difference $(J(u+\varepsilon v)-J(u-\varepsilon v))/(2\varepsilon)$ to the inner product $\la v,\nabla J(u)\ra$. If it converges for sufficiently small $\varepsilon$, the gradient is indeed a viable search direction. If not, there is an error in the implementation of the optimality system. At the same time, this procedure is a test for the accuracy of the numerical scheme.

For the OCT results within MCTDHB, the function space discussed in Sec.~\ref{optsysgen} turned out to be of importance. Optimization within $L^2$ yields controls related to Josephson optimization, whereas within $H^1$ we recover Fock optimization. Within $L^2$ predominantly the local structure of the control is modified, and oscillations are added. These also lead to condensate oscillations, which are hard to stop at the end of the control sequence and thus conflict with the trapping term in the cost function. Therefore, the choices of the best initial control fields and the ratio $\gamma_1/\gamma_2$ turned out to be crucial. In contrast, the $H^1$ optimization is capable of changing the global structure of the control, and is therefore less susceptible to such details.

\subsection{Limitations of a two-parameter optimization}

\begin{figure}
 \begin{tabular}{l}
 (a)\\
 \includegraphics[width=0.72\columnwidth]{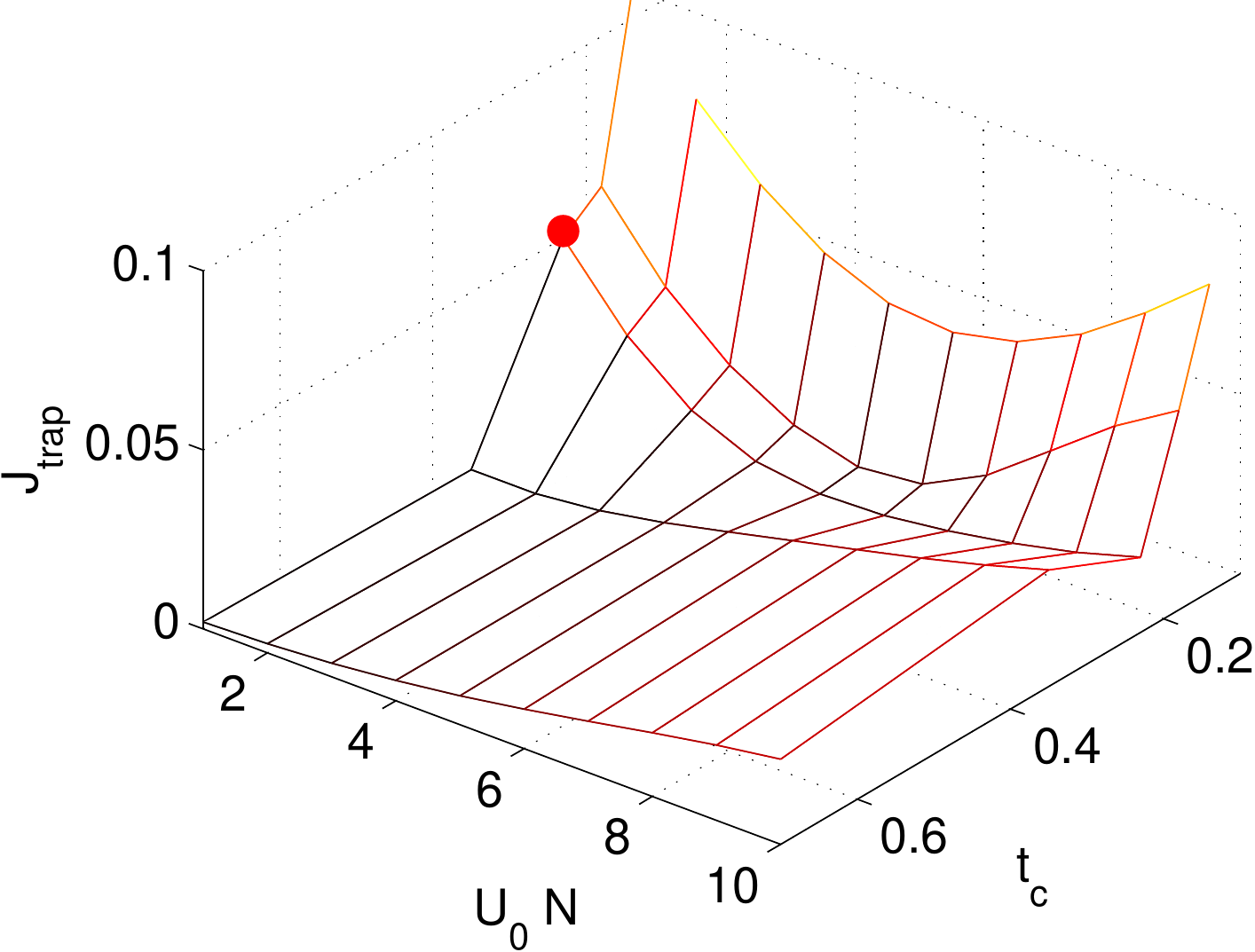}\\
 (b)\\
 \hspace{5mm}\includegraphics[width=0.7\columnwidth]{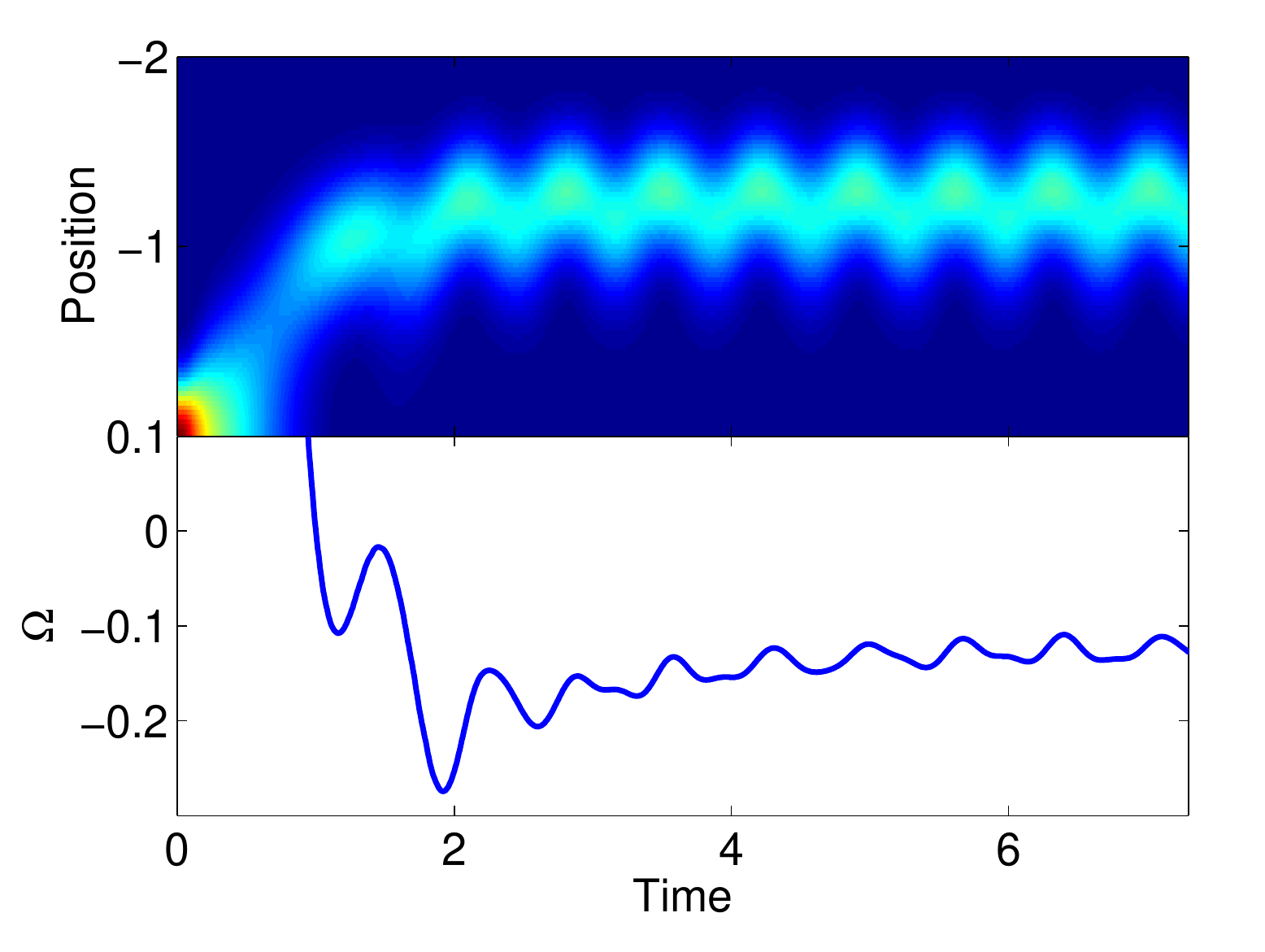}
 \end{tabular}
\caption{ (Color online) (a) Trapping cost function for a two-parameter control for several $U_0 N$ and $t_c$. (b) Example for the density and the tunnel coupling for a specific example ($U_0 N=1$, $t_c=0.23$, marked by the red point), which both strongly oscillate. Here we have $N=100$, but results are similar for different $N$. \label{fig:8}}
\end{figure}

In principle, a two-parameter optimization, as presented in section~\ref{sec:Gen}, is also possible within the realistic model. However, the splitting time scale $t_c$ cannot be made arbitrarily small due to the spatial orbital dynamics. For a too fast splitting, the condensates and thus also the tunnel coupling strongly oscillate. Although we found that small oscillations in $\Omega$, which are always present, do not disturb much, strong oscillations typically prohibit large squeezing. In addition, too strong condensate oscillations during splitting lead to unwanted excitations and to the breakdown of the two-mode approximation. 

In order to quantify those oscillations we calculate the cost functional for trapping from Eq.~\eqref{eq:costMCTDHB}  for different $t_c$ and $U_0$ (the value of $\Omega_c$ does not matter for the present concerns), see Fig.~\ref{fig:8}(a). In panel (b) we report the atomic density and the tunnel coupling for a specific example. At these parameters the cost functional is larger than say $0.025$, $\Omega$ strongly oscillates, and no significant squeezing can be achieved.

From this we conclude that the two-parameter optimization works only for large enough splitting times, which is restrictive primarily for stronger interactions and larger particle numbers. The green lines in Fig.~\ref{fig:3} report the shortest times on which the two-parameter control works within MCTDHB, where the shortest allowed time scale $t_c$ for each $U_0 N$ is taken from Fig.~\ref{fig:8} and two-parameter optimization performed within the generic model to get an estimate. For easier comparison, all results in Fig.~\ref{fig:3} are rescaled such that the adiabatic squeezing time scale ($1/U_0 N$) is the same. 

Decoupling the condensates is necessary after splitting to avoid any residual tunnel coupling. Thus we separate the split condensates further after high squeezing has been achieved, say to a value of $\lambda=1.2$, which corresponds to a double well separation of about 3 $\mu$m. During this separation the number fluctuations might still change, which makes a simple control strategy like the two-parameter optimization from Sec.~\ref{sec:Gen} rather problematic. This can be easily accounted for within OCT as we discuss below.

\subsection{Optimization and results}
\emph{Josephson optimization.} The initial guess is obtained from the tunnel coupling for the ground states shown in Fig.~\ref{fig:7}(c). We choose the ramp such that the tunnel coupling for the static orbitals first quickly drops in an exponential fashion, and add a linear function, similar to our optimization within the generic model. In the first stage, the trap is split quickly up to the point where the tunnel coupling is comparable to the nonlinear interactions. In the second stage, the interplay between tunneling and nonlinear coupling can be exploited to bring the fluctuations down. 

\begin{figure}
 \begin{tabular}{l}
 \includegraphics[width=0.7\columnwidth]{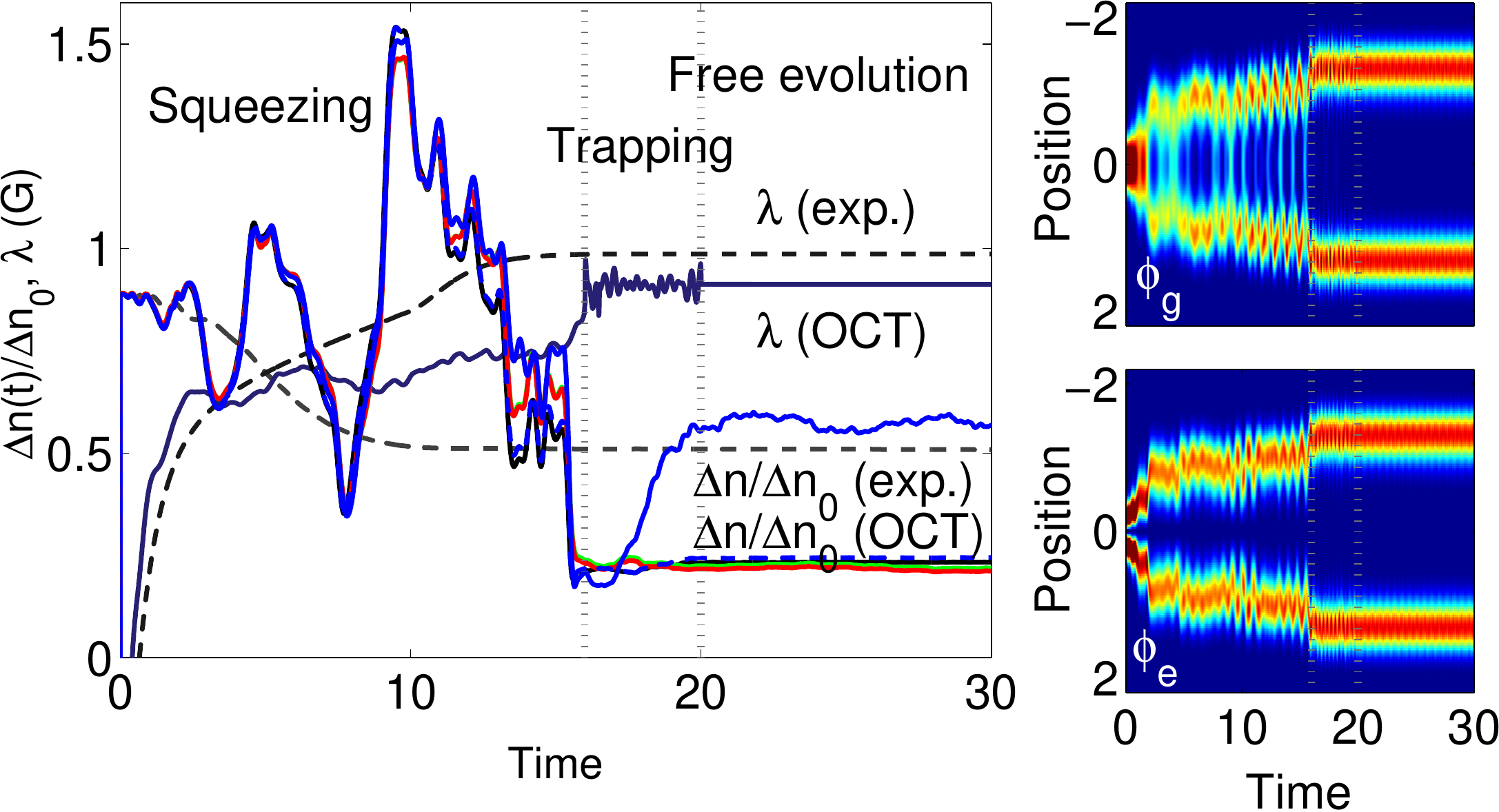}
 \end{tabular}
\caption{ (Color online) In the left panel, a typical solution for \emph{Josephson} optimization for $N=100$. Changing interactions (red line) or atom number (green line) by $10\,\%$ do not appreciably change results. However, the control scheme is sensitive to the (transverse) trap parameters. Here, the deviation of the initial harmonic trapping frequency should not exceed $0.1\,\%$ (dashed blue line) and is crucial for $1\,\%$ (solid blue line). In the right panels, the corresponding orbitals are shown. \label{fig:9}}
\end{figure}

\begin{figure}
\centerline{\includegraphics[width=0.7\columnwidth]{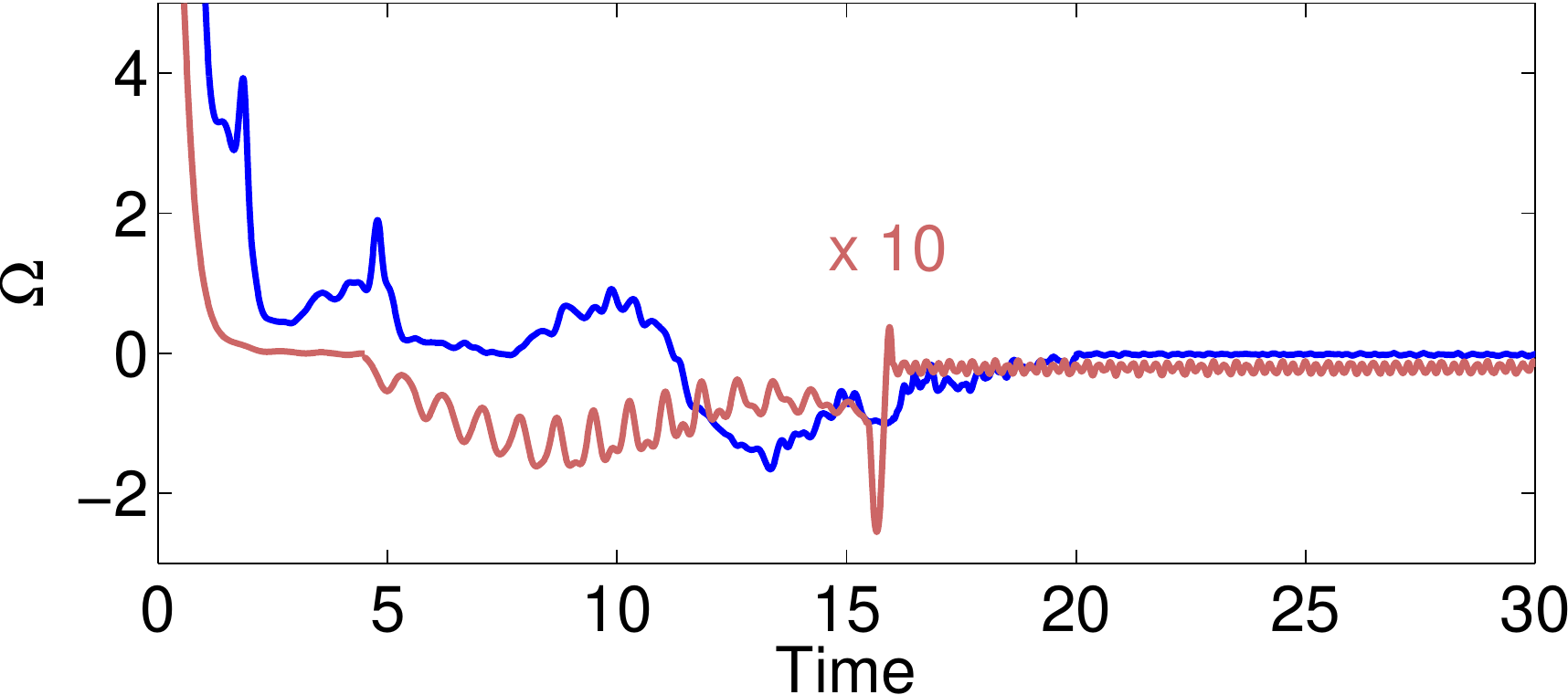}}
\caption{ (Color online) Tunnel coupling within MCTDHB for the \emph{Josephson} (dark blue line) and the \emph{Fock} (bright red line) solution from Fig.~\ref{fig:9} and Fig.~\ref{fig:11} (a), respectively, where the second one is scaled by a factor of $10$.\label{fig:10}}
\end{figure}

Our optimization is performed sequentially, where (i) we optimize squeezing and bring the orbitals to a halt (here both $\gamma_1$ and $\gamma_2$ are finite), and (ii) further separate the condensates and trap them (here only $\gamma_2$ is finite). After these two steps (iii) a waiting period follows (no optimization here), which mimics the stage where the condensate in one well of the interferometer might acquire a phase.
With this initial guess, OCT again comes up with a control reminiscent of parametric amplification, as shown in Fig.~\ref{fig:9}. The large oscillations are due to the parametric oscillations, and the small ones due to the unavoidable oscillations of the orbitals. 
In Fig.~\ref{fig:10} we show the tunnel coupling of the control, which exhibits the oscillatory behavior expected for parametric amplification \cite{grond.praR:09}. When comparing to the generic model, one expects nonlinear renormalizations to $\Omega$ due to the overlap integrals $W_{kqlm}$ \cite{ananikian:06}, which heavily oscillate here. In this sense, the comparison of $\Omega$ in both models can be only a qualitative one and a negative $\Omega$ has no physical significance.  


The parameters for which Josephson optimization works are restricted by the time scale of the squeezing dynamics, determined by $1/U_0 N$, which should be much larger than the time scale determined by the transverse trap frequency. For $U_0 N\gg 1$ the strong oscillations of the orbitals would spoil the control.     
Stability due to uncertainties in the pertinent parameters is also displayed in Fig.~\ref{fig:9}. The control is stable to about $10\%$ variations of the atom number or the nonlinearity. Changes in trap frequency or shape of the double-well, on the other hand, influence the oscillations in the orbitals more significantly and can degrade the control, as discussed in the figure caption.

\begin{figure}
\begin{tabular}{l}
(a)\\
\includegraphics[width=0.7\columnwidth]{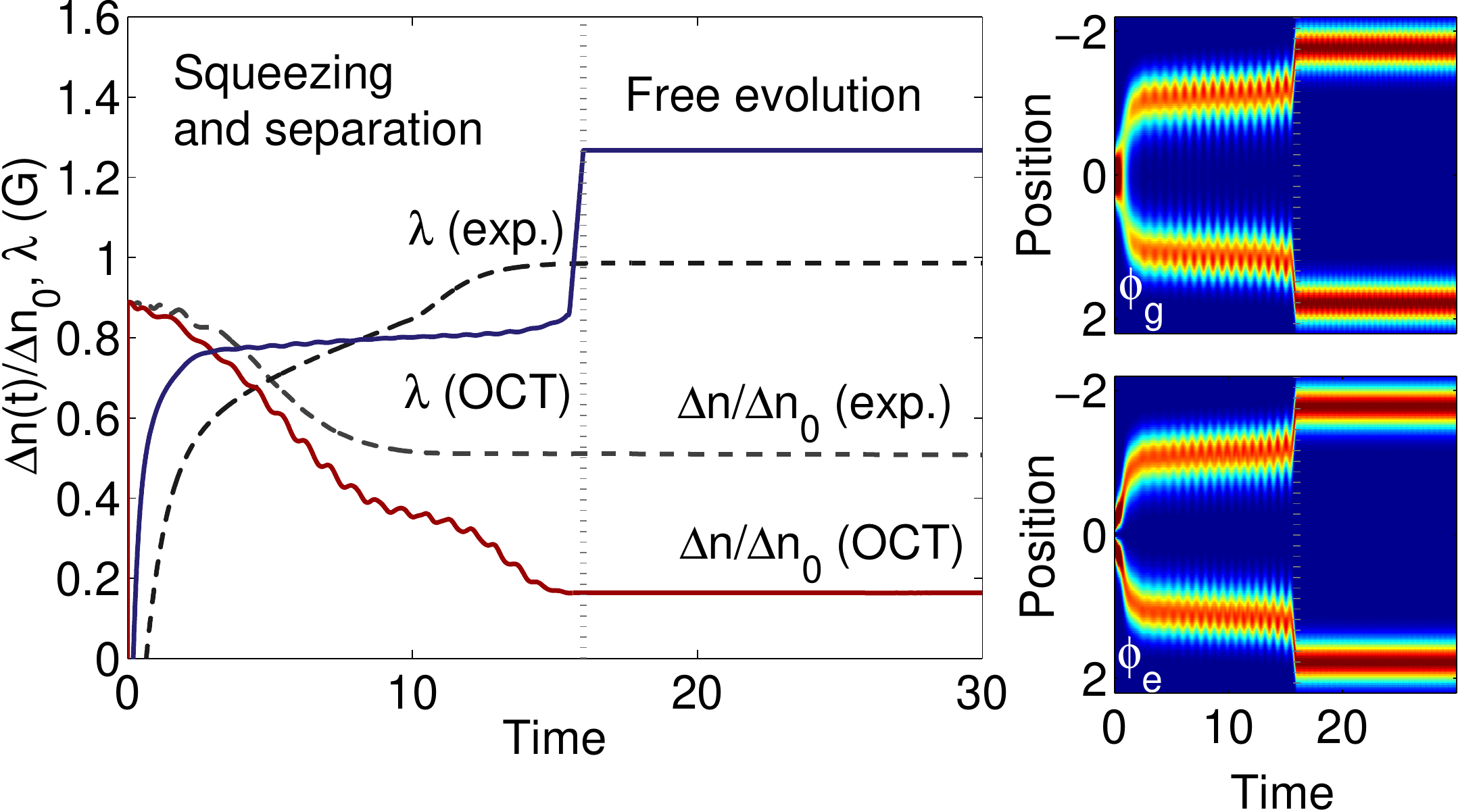}\\
(b)\\
\includegraphics[width=0.7\columnwidth]{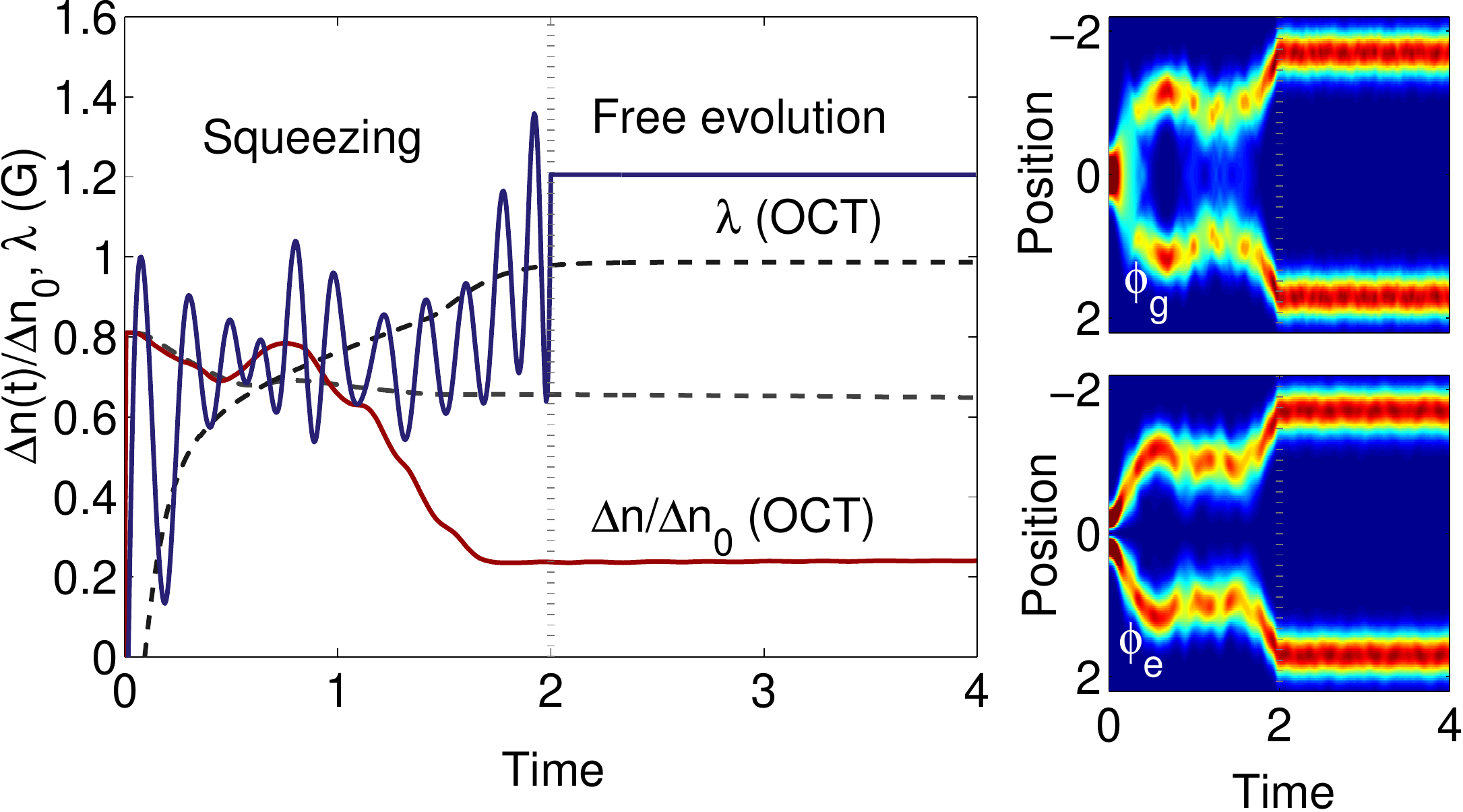}
\end{tabular}
\caption{ (Color online) In the left panels typical OCT solution for \emph{Fock} optimization for $N=100$ with (a) weak interactions $U_0 N=1$ and (b) strong interaction $U_0 N=5$. Exponential (OCT) control and squeezing is displayed by the dashed (solid) lines. In the right panels, the orbitals are shown.  \label{fig:11}}
\end{figure}

\emph{Fock optimization.}---When using the $H^1$ optimization the control becomes smoother. The tunnel coupling depicted in Fig.~\ref{fig:10} shows a behaviour characteristic for Fock optimization. Typical solution are given in Fig.~\ref{fig:11}. In contrast to Josephson optimization, suitable values of $\gamma_1$ and $\gamma_2$ can be found more easily, and the condensates always end up in a stationary state. Here, we demonstrate that we can separate the condensates quite far, while they are at rest at the end of the control sequence. When separating less, the condensates motion during the control sequence is even less oscillatory \cite{hohenester:09}. The sensitivity to the (transverse) trapping potential is much less severe than for Josephson optimization.

Within Fock OCT we find solutions also for stronger interactions ($U_0 N=5$), where the orbitals dynamics is more crucial, as shown in Fig.~\ref{fig:11} (b).

\begin{figure}
  \centerline{\includegraphics[width=0.7\columnwidth]{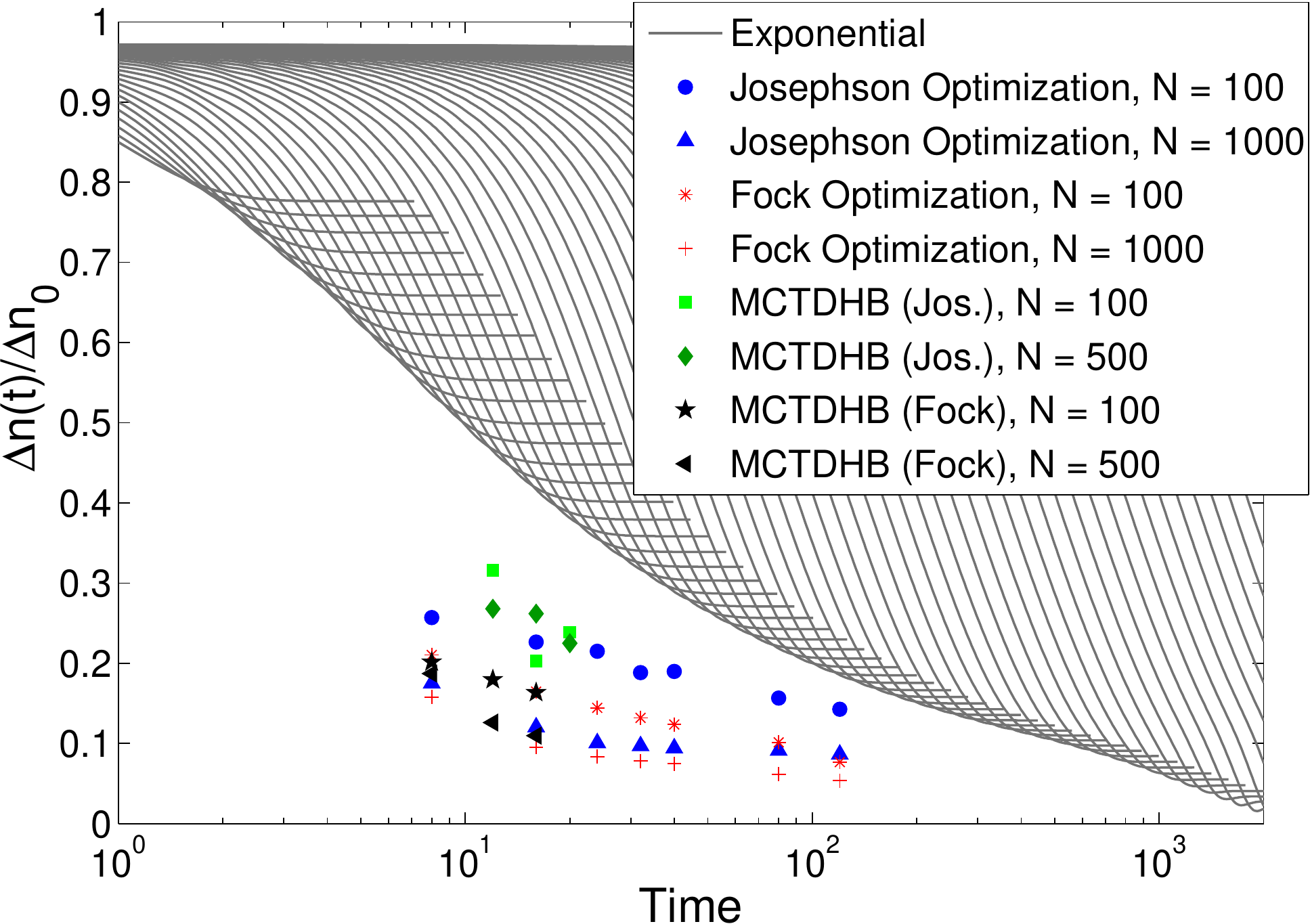}}
\caption{ (Color online) Comparison of squeezing (symbols) within the generic model and MCTDHB for $U_0 N=1$, compared to quasi-adiabatic exponential splitting (gray lines). The latter is calculated within the generic model, since MCTDHB results are very similar. Best results are obtained for \emph{Fock} optimization and large $N$.\label{fig:12}}
\end{figure}

Squeezing values for both OCT strategies in the regime of weak interactions ($U_0 N=1$) are shown in Fig.~\ref{fig:12}, where we also take into account the time it takes to uncouple and trap the condensates. We also present a comparison with the generic model. 

\section{Optimal squeezed states in atom interferometry \label{sec:AI}}

\begin{figure}
 \begin{tabular}{l}
 (a)\\
 \includegraphics[width=0.6\columnwidth]{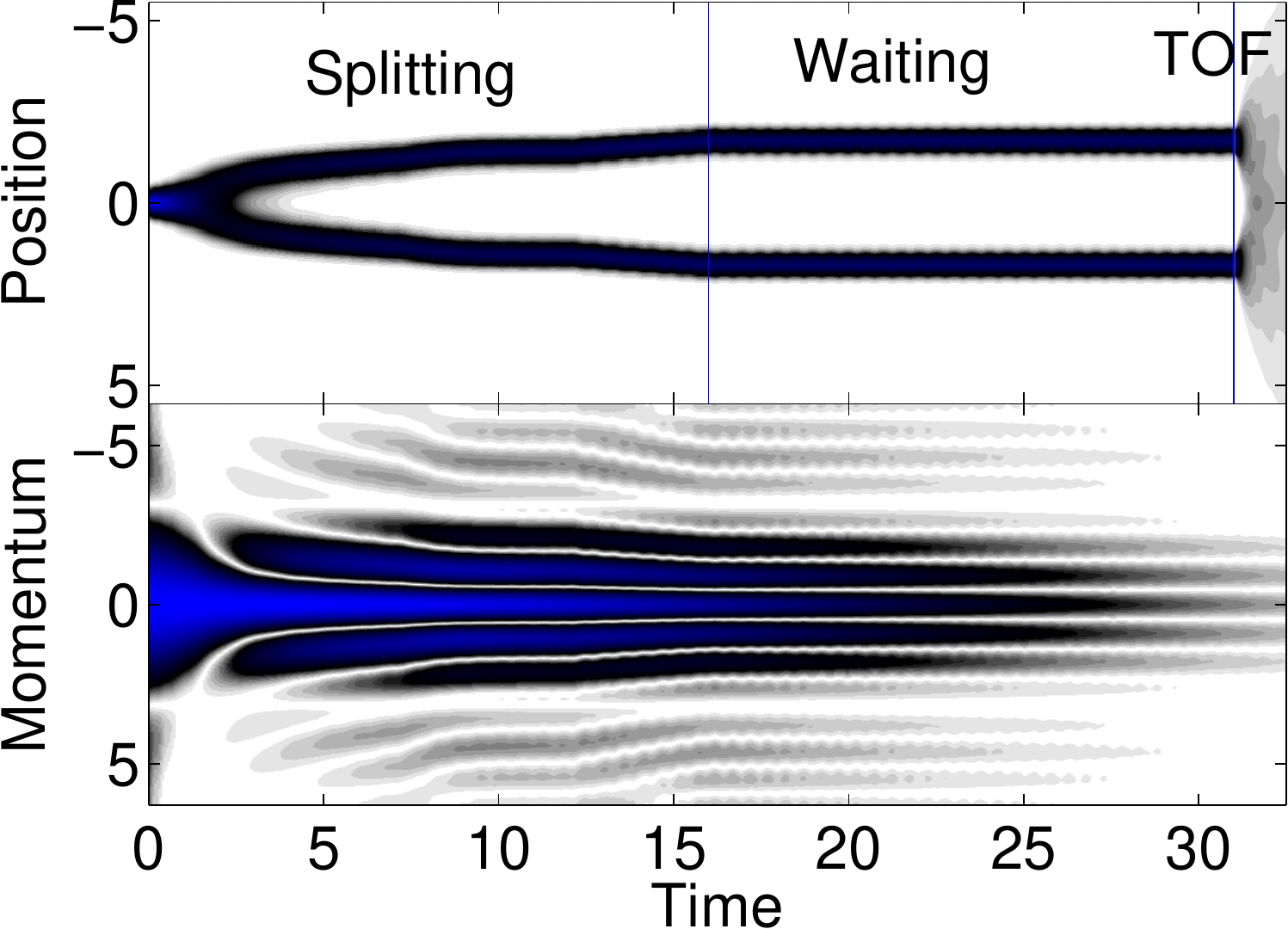}\\
  (b)\\
 \includegraphics[width=0.6\columnwidth]{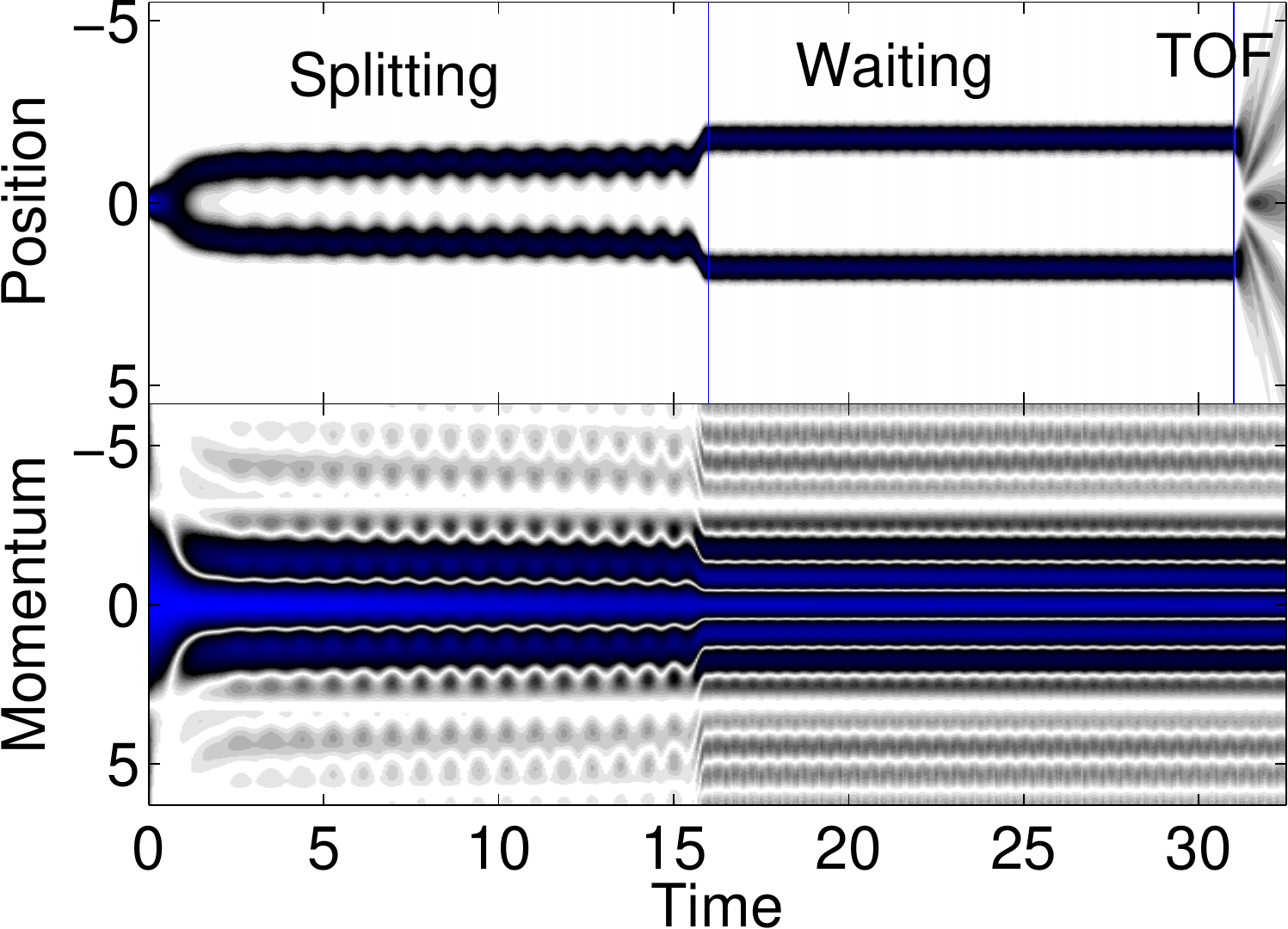}
 \end{tabular}
\caption{ (Color online) In the upper panels, the density for (a) an exponential and (b) an optimized splitting is shown. In the last sequence the trap is switched off and the condensates start to expand and overlap (TOF). In the lower panels, the interference in momentum space, obtained from the Wigner function $W(x=0,p)$ \cite{hohenester:09}, is plotted. When the condensates are released, the interference in momentum space is transformed to an interference in position space. The fringe contrast is reduced in case of exponential splitting due to phase diffusion, and squeezed states are much more robust. For a perfect fringe contrast, the limitations to the phase sensitivity are given by the quantum noise. \label{fig:13}}
\end{figure}

In atom interferometry \cite{cronin:09}, a phase difference between the split condensates is acquired due to the interaction with a weak external potential. For the readout of the phase difference, several schemes exist. In time-of-flight (TOF) experiments the trap is switched off, the condensates expand freely and finally overlap. From the interference fringes the phase can be deduced \cite{shin:04,schumm:05,albiez:05}. The densities for typical control sequences consisting of splitting, waiting (or phase accumulation) time, and recombination in TOF, are shown in the upper panels of Fig.~\ref{fig:13}. 

Another method is phase sensitive recombination of the split condensates, as has been demonstrated in \cite{jo2:07}. Even more desirable would be a Mach-Zehnder (MZ) like setup \cite{pezze:05}, where a $\pi/2$ pulse of the tunnel coupling \cite{pezze2:06} transforms the initial number squeezed state into a phase squeezed state, and after phase accumulation the phase information is transformed into number information by another $\pi/2$-pulse. The readout is achieved then by a relative atom number measurement \cite{esteve:08}.  

Squeezed states have reduced quantum fluctuations and thus phase (number) squeezed states have the potential to increase the phase sensitivity of an atom interferometer below the standard quantum limit $\Delta\phi=1/\sqrt{N}$ \cite{wineland:94,esteve:08} in a TOF (MZ) setup. Phase squeezed states can be obtained from number squeezed states by applying a $\pi/2$ tunneling pulse. The phase sensitivity in terms of shot noise ($1/\sqrt{N}$) is given by the factor of \emph{useful squeezing} $\xi_R=\frac{2\Delta n}{\sqrt{N}\alpha}$ \cite{pezze:05,jaeaeskelaeinen:04}, where $\alpha$ is the coherence factor \cite{pitaevskii:01}. It determines the fringe contrast seen in TOF experiments and is given by the polarization of the state along the $x$-direction of the Bloch sphere $\alpha=\frac{2\langle \hat J_x\rangle}{N}= (\rho_{gg}-\rho_{ee})/N$. In principle, its value tends to zero for very small number fluctuations due to the uncertainty relation between number and phase [see Fig.~7 (a)], but for the squeezed states of our concern here, the limiting factor is the phase diffusion to be discussed below. $\xi_R$ is bounded from below by the fundamental Heisenberg limit $\xi_R=\sqrt{2/N}$~\cite{giovannetti:04}. 
\begin{figure}
 \begin{tabular}{l}
 \includegraphics[width=0.7\columnwidth]{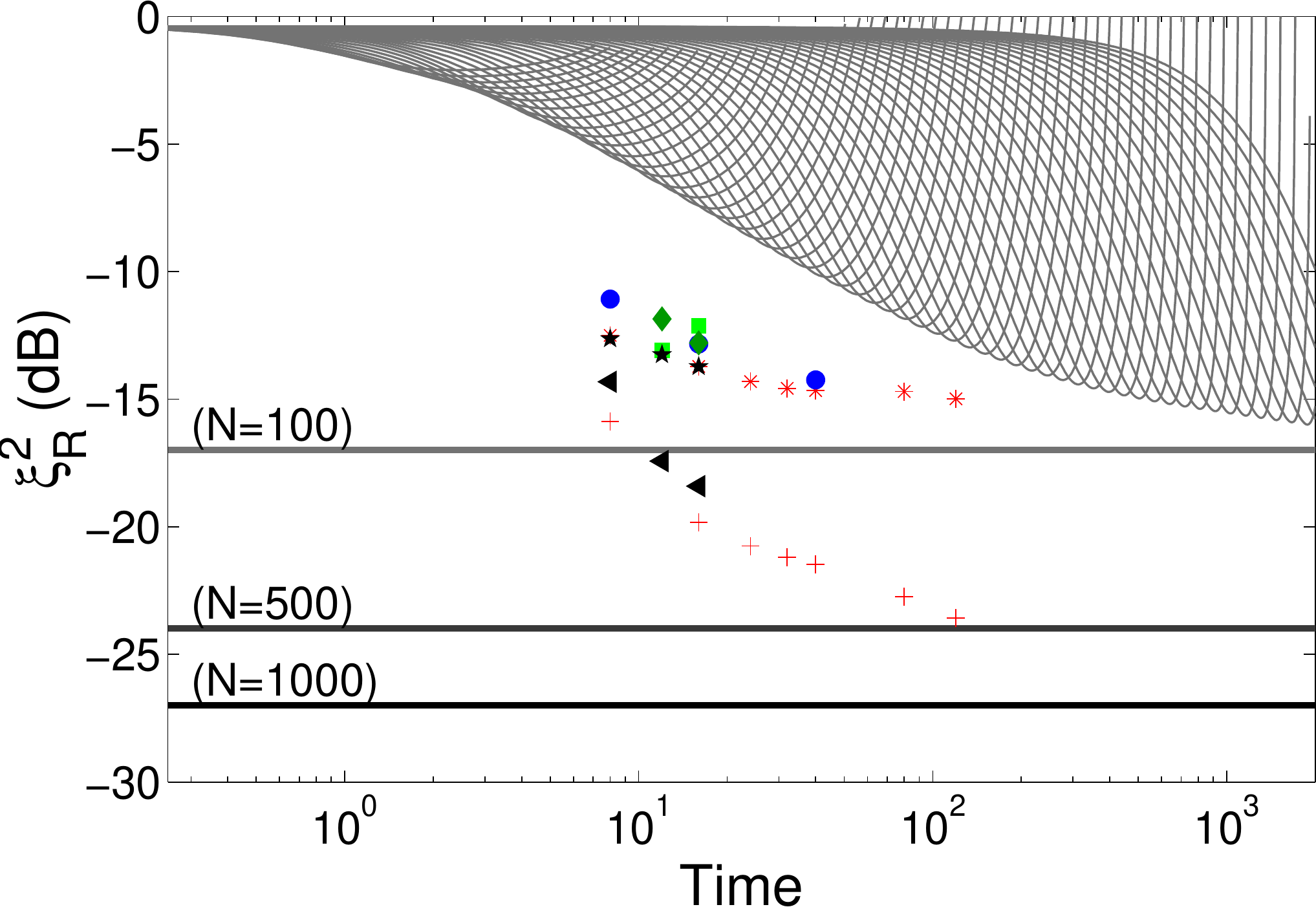}
 \end{tabular}
\caption{ (Color online) $\xi_R$ for optimized (symbols) and exponential splitting (gray lines). The quasi-adiabatic exponential splitting is shown for several time scales for $N=100$. After it reaches a minimum, it rises again due to loss of phase coherence. With optimized splitting, lower $\xi_R$ is obtained at shorter time scales. The symbols are the same as in Fig.~\ref{fig:12}. The results are compared to the Heisenberg limit $\xi_R=\sqrt{\frac{2}{N}}$ (horizontal lines). For larger atom number $N=500$ (MCTDHB) and $N=1000$ (generic model) optimized splitting yields lower $\xi_R$, whereas for adiabatic splitting the curves are similar.  
\label{fig:14}}
\end{figure}
\begin{figure}
 \begin{tabular}{l}
 \includegraphics[width=0.6\columnwidth]{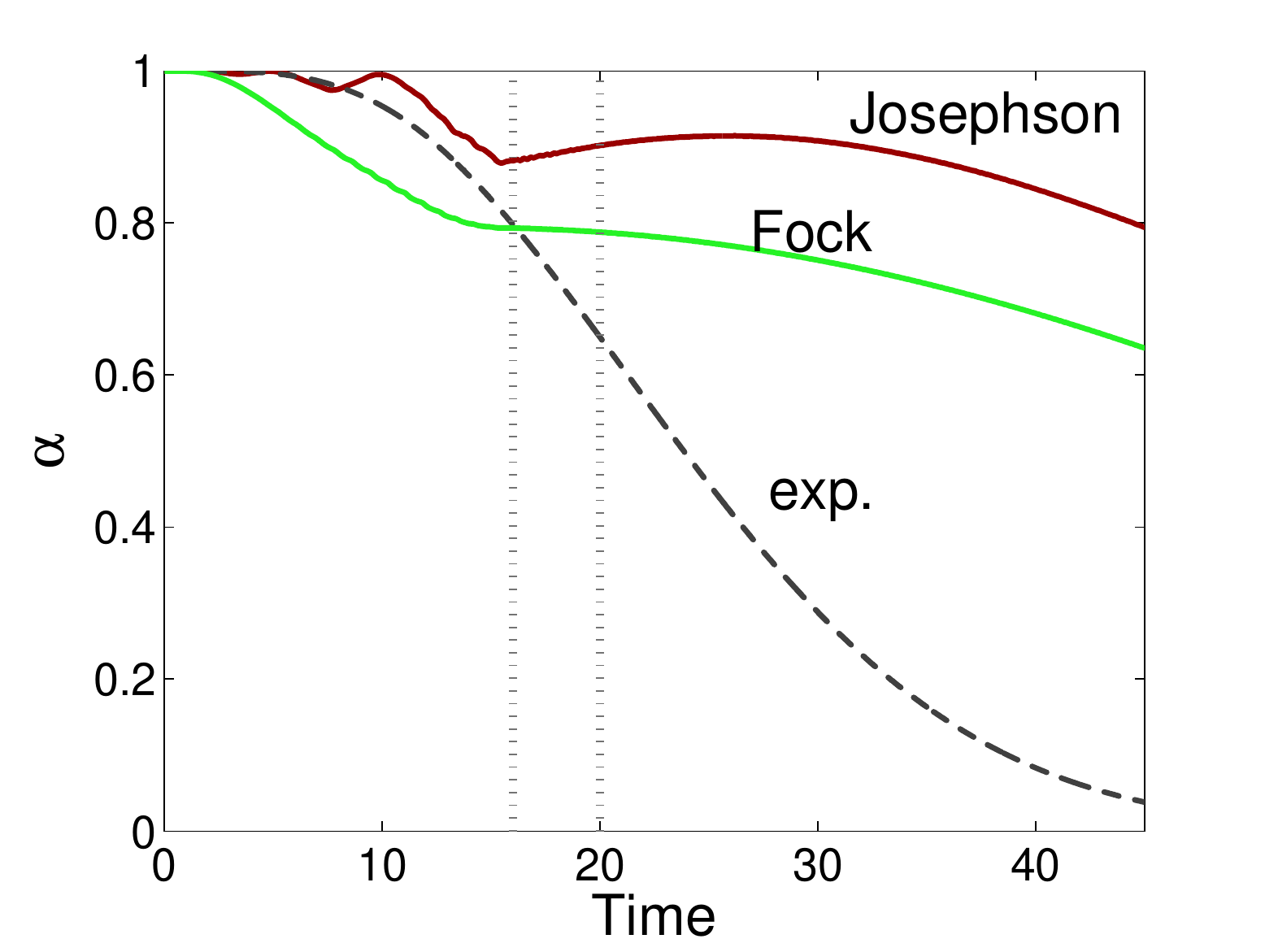}
 \end{tabular}
\caption{ (Color online) Coherence factor $\alpha$ for typical \emph{Josephson} OCT (dark red) and \emph{Fock} OCT (bright green) solutions, where it decays slower than for the quasi-adiabatic exponential splitting (broken line). The Josephson solution shows up oscillations in the phase coherence during splitting, because the driving of the parametric oscillator affects both conjugate observables of number and phase. \label{fig:15}}
\end{figure}

In Fig.~\ref{fig:14} we plot $\xi_R$ for the optimal and the quasi-adiabatic exponential splitting, demonstrating the improvement gained by applying OCT. Note that for the exponential control the minimum of $\xi_R$ is reached at finite values of $\Omega$, where the squeezing still changes in time. Finite tunnel coupling $\Omega$ leads to phase locking between the condensates \cite{albiez:05,hofferberth:07}, which complicates measurement protocols. Thus, additional control strategies would be necessary to uncouple the condensates at the right time.  Josephson and Fock OCT results yield quite similar $\xi_R$, since in the first case there is less squeezing but better phase coherence. 

In presence of interactions, the condensates are subject to phase diffusion \cite{javanainen:97} which quickly reduces the phase sensitivity. On the Bloch sphere, the states are twisted around the $z$-axis, which leads to a decrease of the phase coherence. Whereas phase diffusion, which is proportional to $\Delta n$ \cite{javanainen:97}, is very rapid for binomial or phase squeezed states, number squeezed states are much more robust. However, using number squeezed states in a TOF interferometer comes at the price of a phase sensitivity above shot noise. In Fig.~\ref{fig:15} we compare the phase coherence for various splitting protocols during and after splitting.  Obviously, both OCT results are much more stable against phase diffusion as compared to an exponential splitting on the same time scale. This leads to an improved fringe contrast, as is shown in the lower panels of Fig.~\ref{fig:13} using the Wigner function. Thus, squeezed states allow for much longer phase accumulation times.\footnote{The question which atom number is ideal for reducing phase diffusion in the traps under consideration can be answered as follows. The phase diffusion rate \cite{javanainen:99} is determined by $\sim U_0\Delta n $. From Fig.~\ref{fig:6} we infer that larger $N$ means less interactions $U_0$, and in addition the relative squeezing achieved within OCT is better for larger $N$ (in contrast to the absolute squeezing $\Delta n$). Finally, it turns out that the phase diffusion rate is roughly the same for all $N$ (when using OCT splitting). However, higher number squeezing means enhanced phase fluctuations.}

\section*{Conclusions and Summary}

We discussed optimal control of the splitting of a Bose-Einstein condensate such that number squeezing is minimized. For a physically realistic modeling of the splitting, we employed the MCTDHB equations where the spatial dynamics is included and the control is through the trapping potential. This allows for a direct application of our splitting protocols to experiments. Optimal control of MCTDHB turned out to be sensitive to different mathematical strategies, which give rise to different physical solutions. The optimization of a generic two-mode model allowed us to obtain an intuitive understanding of the underlying physical processes. Two control strategies have been identified, which were compared to a more simple two-parameter optimization. The latter, however, has only limited validity in case of realistic splitting. Results have been given for several squeezing time scales and interactions strength, demonstrating that with refined protocols number squeezing is achieved at least one order of magnitude faster than with exponential splitting. Finally, we analyzed the squeezed states in context of atom interferometry.

\section*{Acknowledgements}

We thank Alfio Borz\`\i, Ofir Alon, Thorsten Schumm, and Armin Scrinzi for most helpful discussions. This work has been supported in part by the Austrian Science Fund FWF under project P18136--N13.

\begin{appendix}

\section{Derivation of the parametric oscillator model \label{app:A}}

In this appendix we present details on how to derive the resonance frequency for parametric amplification and Eq.~\eqref{eq:env}. We start by considering the Hamiltonian of Eq.~\eqref{eq:twomodeosc}, which for $\kappa=0$ is a harmonic oscillator with a time dependent mass $m(t)=\frac{2}{\Omega(t)N}$ and frequency $\omega(t)=\Omega(t)$. For the further steps it helps to rewrite the Hamiltonian using normal variables, which read \cite{messiah:65}
\be  
\hat a=\frac{1}{\sqrt{N}}(k+\frac N2\frac{\partial}{\partial x})\,,\quad \hat{a}^{\dagger}=\frac{1}{\sqrt{N}}(k-\frac N2\frac{\partial}{\partial x})\,.
\ee
We obtain
\begin{equation}\label{eq:twomodeoscTF}
  i\dot C=\tilde{\Omega}(t)\bigl\{(\hat{a}^{\dagger}\hat{a}+\frac 12) + \frac{\kappa N}{2\tilde{\Omega}}(\hat{a}^2+\hat{a}^{\dagger 2})\bigr\}C\,,
\end{equation}
with $\tilde{\Omega}(t)=\Omega(t)+\kappa N$. In order to get rid of the time dependence in the noninteracting part, we transform to the variable 
\be \label{eq:soft} 
s(t)=\int_0^t dt'\tilde{\Omega}(t')\,.
\ee
We are now left with an harmonic oscillator which is driven nonlinearly, a model which is frequently used, e.g., in quantum optics \cite{scully:97}. In an interaction representation with respect to $\hat H_0=\hat{a}^{\dagger}\hat{a}+\frac12$ the Hamiltonian reads
\be\label{eq:twomodeosc2}
\hat H=\frac{\kappa N}{2\tilde{\Omega}(t(s))}(\hat a^2e^{-2is}+(\hat a^{\dagger})^2e^{2is})\,.
\ee
In the spirit of our OCT control (see Fig.~\ref{fig:4}) we assume that the tunnel coupling is an oscillating function with some constant (or slowly varying) offset. The resonance condition is fulfilled for $\tilde{\Omega}(t(s))=\tilde{\Omega}_0+\Omega_1\cos2s$, where the dependence of $t$ on $s$ for this case can be obtained from Eq.~\eqref{eq:soft}:
\be
t(s)=\frac{1}{\tilde{\Omega}_0}\Bigl\{\frac{1}{\sqrt{1-\beta^2}}\arctan{\Bigl[\frac{(1-\beta)\tan{s}}{\sqrt{1-\beta^2}}\Bigr]}\Bigr\}\,.
\ee
Assuming $\beta=\frac{\Omega_1}{\tilde{\Omega}_0}\ll 1$, we have to a good approximation simply $s=\tilde{\Omega}_0 t$. This gives us the resonance frequency $\omega_{res}=\tilde{\Omega}_0$.

For discussing the time scale of the decrease of the number fluctuations, we separate the Hamiltonian of Eq.~\eqref{eq:twomodeosc2} into a resonant and a non-resonant part. Using $\frac{1}{1+\beta \cos{(2\omega_{res}t)}}\approx 1-\beta\cos{(2\omega_{res}t)}$, the resonant part reads in a rotating wave approximation \cite{scully:97}
\begin{equation}\label{eq:twomodeosc3}
  \hat H_I^{res}=-\frac{\kappa N\Omega_1}{4\tilde{\Omega}_0^2}
  \left(\hat a^2+  \hat a^{\dagger\,2}\right)\,.
\end{equation}
Solving for the fluctuations of a general quadrature $k^{\mu}=\frac{\sqrt{N}}{2}(ae^{ i\mu}+a^{\dagger}e^{-i\mu})$ yields
\be
\bigl(\Delta|k_{\mu}|\bigr)^2=\frac N4\Bigl(e^{\frac{\kappa N\Omega_1}{\tilde{\Omega}_0}t}\cos^2(\mu)+e^{-\frac{\kappa N\Omega_1}{\tilde{\Omega}_0}t}\sin^2(\mu)\Bigr)\,.
\ee
Thus, one of the quadrature observables (here $\mu=\frac{\pi}{2}$) gets squeezed, while another quadrature (here $\mu=0$) gets anti-squeezed. Transforming back to the Heisenberg picture amounts to (assuming again $\beta\ll 1$) adding a time dependent phase  $\mu\rightarrow\mu+\tilde{\Omega}_0 t$. As a result the parametric oscillations lead to squeezing in a quadrature which rotates. This explains the oscillations (see Fig.~\ref{fig:4}) in the fluctuations of the observable we are interested in, namely $\Delta n=\Delta |k_{\mu}|$, and gives rise to the expected decay of the envelope in Eq.~\eqref{eq:env}.

\section{Optimality system for MCTDHB \label{app:B}}

In this appendix we derive for the MCTDHB method the control equation, the terminal conditions for the adjoint variables, and the equations of motion (eom) for the adjoint variables. The OCT approach is the same as in Sec.~\ref{sec:Gen}, where the Fr\'echet derivatives with respect to the variables $\phi_g$, $\phi_e$, $\mathbf{C}$, the adjoint variables $\tilde\phi_g$, $\tilde\phi_e$, $\tilde{\mathbf{C}}$, and the control $\lambda$ are equated to zero. 

The Lagrange functional is given analogous to Eq.~\eqref{eq:Lag}:
\bea\label{eq:LagApp}
&& L(\phi_g,\phi_e,\mathbf{C},\tilde{\phi}_g,\tilde{\phi}_e,\tilde{\mathbf{C}},\lambda) =  J(\phi_g,\phi_e,\mathbf{C},\lambda)+\\
&& \sum_k Re\langle \tilde{\phi}_k, i\dot{\phi_k}-\hat{\mathbf{P}}\bigl[\hat{h}\phi_k+U_0 \{\mathbf{\rho}\}_{kk}^{-1}\sideset{}{'}\sum_{s,q,l}\rho_{ksql}\phi_s^*\phi_q\phi_l\bigr]\rangle\non
&& + Re\langle \tilde{\mathbf{C}}, i\dot{\mathbf{C}}   -\bigl(-\Omega\hat{J}_x+\frac{1}{2}\sideset{}{'}\sum_{k,q,l,m}\hat a_k^{\dagger}\hat a_q^{\dagger}\hat a_l\hat a_m W_{kqlm} \bigr)\mathbf C \rangle\,,\nonumber
\eea
where the brackets again imply integrating over time. The cost functional $J(\phi_g,\phi_e,\mathbf{C},\lambda)$ is given in Eq.~\eqref{eq:costMCTDHB}.

The control equation reads
\bea
\nabla J&=&-\gamma\ddot{\lambda}-\sum_k Re\langle \tilde{\phi}_k|\frac{\partial V_{\lambda}}{\partial\lambda}|\phi_k\rangle\\
&&+Re\sum_{q,k}\langle \tilde{\phi}_q|\phi_k\rangle\langle\phi_k|\frac{\partial V_{\lambda}}{\partial\lambda}|\phi_q\rangle\non
&&-Re\langle\tilde{\mathbf{C}}|\hat{J}_x|\mathbf{C}\rangle \sum_k (-1)^k\langle\phi_k|\frac{\partial V_{\lambda}}{\partial\lambda}|\phi_k\rangle\,,\nonumber
\eea
where $(-1)^g=1$ and $(-1)^e=-1$. In case of $H^1$ formulation we get a Poisson equation for the gradient, with the rhs from above.

When calculating the Fr\'echet derivatives, partial time integration has to be performed, which yields the terminal conditions for the adjoint variables. Here it is necessary to make the symmetric definition $d=\bigl(\int_{-\infty}^{0} dx\phi_g^*[x]\phi_e[x] -\int_{0}^{\infty} dx\phi_g^*[x]\phi_e[x]\bigr)/2$, cf. section~\ref{subsubsec:ground}. We then have
\bea
&&i \tilde\phi_k(T)=-\gamma_2\la\phi_k^d|\phi_k(T)\ra\phi_k^d\\
&&+\gamma_1\langle \mathbf{C}(T)|\bigl\{\frac{\Delta\hat{N}}{2},\hat{a}_k^{\dagger}\hat{a}_q\bigr\}|\mathbf{C}(T)\rangle\frac{\Theta(-x)-\Theta(x)}{2}\phi_q(T)\,,\nonumber
\eea
with $k\neq q$, and
\bea
&&i|\tilde{\mathbf{C}}(T)\rangle=\gamma_1\frac{\Delta\hat{N}^2}{4}|\mathbf{C}(T)\rangle\,.
\eea


When deriving the adjoint equations, we first look at simplified eom and then make the equations subsequently more complete. We start with two simple Schr\"odinger equations with projectors, than we add the nonlinearities, and finally we take into account also the number distribution dynamics.

\subsubsection*{Two Schr\"odinger equations with projectors}

We consider the eom $i\dot{\phi}_i=\hat{\mathbf{P}}\hat{h}\phi_i$ with Lagrangian (we omit arguments for simplicity)
\be
L=J+\sum_i\int_0^T dt\hat{L}\,,\quad\mathrm{with}\quad \hat L=Re\la \tilde{\phi}_i|i\dot{\phi}_i-\hat{\mathbf{P}}\hat{h}\phi_i\ra\;.
\ee
We then need to calculate the rhs of $i\dot{\tilde{\phi}}_i=-2\frac{\partial \hat{L}}{\partial \phi_i^*}$. Without the projectors we would be left with Schr\"odinger equations for $\tilde{\phi}_i$, but here we have in addition the terms
\bea
S_i=
 \sum_k\Bigl([\la\phi_k| \tilde{\phi}_i\ra+\la  \tilde{\phi}_k|\phi_i\ra]\hat{h}\phi_k+\la\phi_k|\hat{h}|\phi_i\ra  \tilde{\phi}_k\Bigr)\,.\nonumber
\eea

\subsubsection*{Two nonlinear Schr\"odinger equations with projectors}
Here we consider the Eqs. \eqref{eq:LagApp}, but with constant coefficients. The Lagrangian is then given by the first two lines of Eq.~\eqref{eq:LagApp}, assuming constant one- and two body reduced densities. Without projectors the relevant terms (to be multiplied by the corresponding coefficients) are
\be
g^i_{lrst}= \tilde{\phi}_l^*\phi_s\phi_t\delta_{ir}\,,
\ee
and
\be
\tilde{g}^i_{lrst}= \tilde{\phi}_l\phi_r(\phi_s^*\delta_{it}+\phi_t^*\delta_{is})\,.
\ee
The projector terms yield
\be
G^i_{lrst}=\la  \tilde{\phi}_l|\phi_i\ra\phi_r^*\phi_s\phi_t+\sum_k\la  \tilde{\phi}_l|\phi_k\ra\phi_k^*\phi_s\phi_t\delta_{ir}\,,
\ee
and
\be
\tilde{G}^i_{lrst}= \tilde{\phi}_l\la \phi_r^*\phi_s \phi_t|\phi_i\ra + \sum_k\la\phi_k|\tilde{\phi}_l\ra\phi_r\phi_k(\phi_t^*\delta_{is} + \phi_s^*\delta_{it})\,.
\ee

\subsubsection*{MCTDHB equations}
Now we consider the complete MCTDHB equations, with Lagrangian given in Eq.~\eqref{eq:LagApp}. For the eom for $ \tilde{\phi}_i$ it requires also to consider derivations of the number distribution part in the Lagrangian, i.e., of the tunnel coupling and the two-body matrix elements. This yields the contributions
\bea
M_i&=&(-1)^i2Re\langle\tilde{\mathbf{C}}|\hat{J}_x| \mathbf{C}\rangle\hat{h}\phi_i\non
&+&U_0/2\sideset{}{'}\sum_{k,q,l,m} (\langle\tilde{\mathbf{C}}|\hat a_k^{\dagger}\hat a_q^{\dagger}\hat a_l\hat a_m|\mathbf{C}\ra+
\langle\mathbf{C}|\hat a_k^{\dagger}\hat a_q^{\dagger}\hat a_l\hat a_m|\tilde{\mathbf{C}}\ra)\times\non
&&(\phi_q^*\delta_{ik}+\phi_k^*\delta_{iq})\phi_l\phi_m
\eea
Using also the previous results we obtain as eom
\bea
&&i \dot{ \tilde{\phi}}_i=\hat{h} \tilde{\phi}_i - S_i\non
&&+ \sum_{k\neq j}U_0\{\mathbf{\rho}\}_{kk}^{-1}\Bigl\{\rho_{kkkk}\bigl(g^i_{kkkk}+\tilde{g}^i_{kkkk}-G^i_{kkkk}-\tilde{G}^i_{kkkk})\non
&&+2\rho_{kjkj}(g^i_{kjkj}+\tilde{g}^i_{kjkj}-G^i_{kjkj}-\tilde{G}^i_{kjkj})\\
&&+\rho_{kkjj}(g^i_{kkjj}-G^i_{kkjj})+\rho_{kkjj}^*(\tilde{g}^i_{kkjj}-\tilde{G}^i_{kkjj})\Bigr\}+M_i\,.\nonumber
\eea
The eom for $\tilde{\mathbf{C}}$ is, besides a two-mode Hamiltonian contribution as in Eq.~\eqref{eq:TMMCham}, given by the derivatives of the densities in the orbital part of the Lagrangian as
\bea
&&i |\dot{\tilde{\mathbf{C}}}\rangle=\mathcal{H}|\tilde{\mathbf{C}}\rangle\\
&&+U_0\sum_{i\neq j}\Bigl\{2 Re\la \tilde{\phi}_i|\mathbf{\hat{P}}|\phi_i|^2\phi_i\ra\bigl(\rho_{ii}^{-1}\hat a_i^{\dagger}\hat a_i \hat a_i^{\dagger} \hat a_i-\frac{\hat a_i^{\dagger} \hat a_i}{\rho_{ii}^2}\rho_{iiii}\bigr)\non
&&+4 Re\la \tilde{\phi}_i|\mathbf{\hat{P}}|\phi_j|^2\phi_i\ra\bigl(\rho_{ii}^{-1}\hat a_i^{\dagger}\hat a_j^{\dagger} \hat a_i \hat a_j-\frac{\hat a_i^{\dagger} \hat a_i}{\rho_{ii}^2}\rho_{ijij}\bigr)\non
&&+2 Re\bigl[\la \tilde{\phi}_i|\mathbf{\hat{P}}\phi_i^*\phi_j^2\ra\bigl(\rho_{ii}^{-1}\hat a_i^{\dagger}\hat a_i^{\dagger} \hat a_j \hat a_j-\frac{\hat a_i^{\dagger} \hat a_i}{\rho_{ii}^2}\rho_{iijj}\bigr)\bigr]\Bigr\}|\mathbf{C}\rangle\,.\nonumber
\eea

\section{Details on the numerical implementation \label{app:C}}

In this appendix we give details on our numerical implementation of the time evolution and optimal control of the MCTDHB equations. Given the initial value problem $\dot{y}=f(y,t)$, \quad $y(0)=y_0$, we employ a Modified Crank-Nicolson time-stepping method \cite{Thomee:97} 
\be
\frac{y^{n+1}-y^n}{\Delta t}=\frac{f(\frac{y^{n+1}+y^n}{2},\frac{t^{n+1}+t^n}{2})}{2}+\mathcal{O}(\Delta t^3)\,,
\ee
which is a norm preserving scheme. At each time step, we solve this equation using Newton iterations, where $y^n$ converges quadratically and we employ the stopping criterion $|y^n-y^{n+1}|<10^{-5}$. This time-stepping scheme is very stable, even when replacing $y^{n+1}\rightarrow y^n$ on the rhs (with loss of accuracy off course).

For Newton's method we have to take functional derivatives of our eom, and due to the projector terms we obtain equations of type
$(A+uv^{\dagger})x=b$, $Ax+(x^{\dagger} v)u=b$, and $Ax+\Re{(v^{\dagger}x)}u=b$, to be solved for the vector $x$. Here $A$ is a sparse matrix and can thus be inverted efficiently, and $u$ and $v$ are vectors. The first equation is linear in $x$ and the term $uv^*$, which arises due to the projectors is a low rank dense matrix, and its inversion is thus computationally very expensive. A way around the direct numerical inversion is the Sherman-Morrison formula \cite{Nocedal:99}, which reads
\be
x=A^{-1} b-\frac{v^{\dagger}A^{-1}b}{1+v^{\dagger}A^{-1}u}A^{-1}u\,,
\ee 
and is solvable whenever $A$ is nonsingular and $v^{\dagger}A^{-1} u\neq -1$. For computational efficiency, $A^{-1}$ is never calculated explicitly, but rather the vectors $A^{-1} b$ and $A^{-1} u$. 

The second and third equations to be solved are not linear in $x$ due to complex conjugation. We can attack this problem by writing every component in terms of real and imaginary parts and use the matrix representation of complex numbers
\be 
x=x_r+i x_i\;\rightarrow\;\Bigl(\begin{array}{c c} x_r &-x_i\\x_i &x_r\end{array}\Bigr)\;.
\ee
We then have to solve the systems 
\be
\Bigl[\Bigl(\begin{array}{c c} A_r &-A_i\\A_i &A_r\end{array}\Bigr)+\Bigl(\begin{array}{c c} u_r & -u_i\\u_i&u_r\end{array}\Bigr)\Bigl(\begin{array}{c c} v_r^T & v_i^T\\v_i^T&-v_r^T\end{array}\Bigr)\Bigr]\Bigl(\begin{array}{c} x_r \\x_i\end{array}\Bigr)=\Bigl(\begin{array}{c} b_r \\b_i\end{array}\Bigr)
\ee
and
\be
\Bigl[\Bigl(\begin{array}{c c} A_r &-A_i\\A_i &A_r\end{array}\Bigr)+\Bigl(\begin{array}{c} u_r \\u_i\end{array}\Bigr)(\begin{array}{c c}v_r^T& v_i^T\end{array})\Bigr]\Bigl(\begin{array}{c} x_r \\x_i\end{array}\Bigr)=\Bigl(\begin{array}{c} b_r \\b_i\end{array}\Bigr)\,,
\ee
respectively, by use of the Sherman-Morrison-Woodbury (SMW) \cite{Nocedal:99} formula
\be
(A+UCV)^{-1}=A^{-1}-A^{-1}U(C^{-1}+VA^{-1}U)^{-1}VA^{-1}\,.
\ee
Here, $A$, $U$, $C$ and $V$ are matrices of size $n\times n$, $n\times k$, $k\times k$ and $k\times n$, respectively, with $k\le n$.
 
The complex extensions to the SMW-formula, which are novel, to our knowledge, are also used for the adjoint equations (see appendix~\ref{app:A}), which are linear in the adjoint variable, but again contain full matrices due to the projectors in Eqs.~\eqref{eq:MCTDHBorbs}.

\end{appendix}


\end{document}